\documentclass[10pt,journal]{IEEEtran}
\usepackage{cite,calc}
\usepackage[cmex10]{amsmath}   
\usepackage{amsfonts,amssymb,graphics,psfrag,subfigure,epsfig,graphics}
\usepackage[mathcal]{euscript}  

\newcommand{\secref}[1]{Section~\ref{#1}}
\newcommand{\SNR}{\text{SNR}}
\newcommand{\cgt}{\dagger}
\newcommand{\tr}{\mathrm{T}}

\DeclareMathOperator{\diag}{diag}

\DeclareMathAlphabet{\mathbsf}{OT1}{cmss}{bx}{n}
\DeclareMathAlphabet{\mathssf}{OT1}{cmss}{m}{sl}
\DeclareMathAlphabet{\mathcsf}{OT1}{cmss}{sbc}{n}

\newcommand{\rvs}[1]{\mathssf{#1}}
\newcommand{\rvv}[1]{\mathbsf{#1}}

\newcommand{\svv}[1]{\mathbf{#1}}

\DeclareSymbolFont{bsfletters}{OT1}{cmss}{bx}{n}  
\DeclareSymbolFont{ssfletters}{OT1}{cmss}{m}{n}
\DeclareMathSymbol{\bsfGamma}{0}{bsfletters}{'000}
\DeclareMathSymbol{\ssfGamma}{0}{ssfletters}{'000}
\DeclareMathSymbol{\bsfDelta}{0}{bsfletters}{'001}
\DeclareMathSymbol{\ssfDelta}{0}{ssfletters}{'001}
\DeclareMathSymbol{\bsfTheta}{0}{bsfletters}{'002}
\DeclareMathSymbol{\ssfTheta}{0}{ssfletters}{'002}
\DeclareMathSymbol{\bsfLambda}{0}{bsfletters}{'003}
\DeclareMathSymbol{\ssfLambda}{0}{ssfletters}{'003}
\DeclareMathSymbol{\bsfXi}{0}{bsfletters}{'004}
\DeclareMathSymbol{\ssfXi}{0}{ssfletters}{'004}
\DeclareMathSymbol{\bsfPi}{0}{bsfletters}{'005}
\DeclareMathSymbol{\ssfPi}{0}{ssfletters}{'005}
\DeclareMathSymbol{\bsfSigma}{0}{bsfletters}{'006}
\DeclareMathSymbol{\ssfSigma}{0}{ssfletters}{'006}
\DeclareMathSymbol{\bsfUpsilon}{0}{bsfletters}{'007}
\DeclareMathSymbol{\ssfUpsilon}{0}{ssfletters}{'007}
\DeclareMathSymbol{\bsfPhi}{0}{bsfletters}{'010}
\DeclareMathSymbol{\ssfPhi}{0}{ssfletters}{'010}
\DeclareMathSymbol{\bsfPsi}{0}{bsfletters}{'011}
\DeclareMathSymbol{\ssfPsi}{0}{ssfletters}{'011}
\DeclareMathSymbol{\bsfOmega}{0}{bsfletters}{'012}
\DeclareMathSymbol{\ssfOmega}{0}{ssfletters}{'012}

\begin{document}

\title{Rateless Coding for Gaussian Channels}
\author{Uri~Erez,~\IEEEmembership{Member,~IEEE,}
Mitchell~D.~Trott,~\IEEEmembership{Fellow,~IEEE,}
Gregory~W.~Wornell,~\IEEEmembership{Fellow,~IEEE} \thanks{Manuscript
received August 2007, revised December 2010.  This work was supported in
part by the National Science Foundation under Grant No.~CCF-0515122, 
Draper Laboratory, MITRE Corp., and by Hewlett-Packard Co. through the
MIT/HP Alliance.  This work was presented in part at the Information
Theory and Applications Workshop, University of California, San Diego,
Feb.\ 2006, and at the International Symposium on Information Theory,
Seattle, WA, July 2006.} \thanks{U.~Erez is with the Department of
Electrical Engineering - Systems, Tel Aviv University, Ramat Aviv,
69978, Israel (Email: uri@eng.tau.ac.il).} \thanks{M.~D.~Trott is with
Hewlett-Packard Laboratories, Palo Alto, CA, 94304 (Email:
mitchell.trott@hp.com).} \thanks{G.~W.~Wornell is with the Department
of Electrical Engineering and Computer Science, Massachusetts
Institute of Technology, Cambridge, MA 02139 (Email: gww@mit.edu).}}

\maketitle

\begin{abstract}

A rateless code---i.e., a rate-compatible family of codes---has the
property that codewords of the higher rate codes are prefixes of those
of the lower rate ones.  A perfect family of such codes is one in
which each of the codes in the family is capacity-achieving.  We show
by construction that perfect rateless codes with low-complexity
decoding algorithms exist for additive white Gaussian noise channels.
Our construction involves the use of layered encoding and successive
decoding, together with repetition using time-varying layer weights.  As
an illustration of our framework, we design a practical three-rate
code family.  We further construct rich sets of near-perfect rateless
codes within our architecture that require either significantly fewer
layers or lower complexity than their perfect counterparts.
Variations of the basic construction are also developed, including one
for time-varying channels in which there is no a priori stochastic
model.

\end{abstract}


\section{Introduction}

\IEEEPARstart{T}{he} design of effective ``rateless'' codes has
received renewed strong interest in the coding community, motivated by
a number of emerging applications.  Such codes have a long history,
and have gone by various names over time, among them incremental
redundancy codes, rate-compatible punctured codes, hybrid automatic
repeat request (ARQ) type II codes, and static broadcast codes
\cite{Chase,Mand,Hag,Lin,RM,SSV,HKM,JS,ct01,sf00}.  This paper focuses
on the design of such codes for average power limited additive white
Gaussian noise (AWGN) channels.  Specifically, we develop techniques
for mapping standard good single-rate codes for the AWGN channel into
good rateless codes that are efficient, practical, and can operate at
rates of multiple b/s/Hz.  As such, they represent an attractive
alternative to traditional hybrid ARQ solutions for a variety of
wireless and related applications.

More specifically, we show that the successful techniques employed to
construct low-complexity codes for the standard AWGN channel---such as
those arising out of turbo and low-density parity check (LDPC)
codes---can be leveraged to construct rateless codes.  In particular,
we develop an architecture in which a single codebook designed to
operate at a single SNR is used in a straightforward manner to build a
rateless codebook that operates at many SNRs.

The encoding in our architecture exploits three key ingredients:
layering, repetition, and time-varying weighting.  By layering, we
mean the creation of a code by a linear combination of subcodes.  By
repetition, we mean the use of simple linear redundancy.  Finally, by
time-varying weighting, we mean that the (complex) weights in the
linear combinations in each copy are different.  We show that with the
appropriate combination of these ingredients, if the base codes are
capacity-achieving, so will be the resulting rateless code.

In addition to achieving capacity in our architecture, we seek to
ensure that if the base code can be decoded with low complexity,
so can the rateless code.  This is accomplished by
imposing the constraint that the layered encoding be successively
decodable---i.e., that the layers can be decoded one at a time,
treating as yet undecoded layers as noise.

Hence, our main result is the construction of capacity-achieving,
low-complexity rateless codes, i.e., rateless codes constructed from
layering, repetition, and time-varying weighting, that are
successively decodable.

The paper is organized as follows.  In \secref{s:back} we put the
problem in context and summarize related work and approaches.  In
\secref{s:model} we introduce the channel and system model.
In \secref{s:example} we motivate and illustrate our construction with
a simple special-case example.  In \secref{s:ldr} we develop our
general construction and show that within it exist perfect rateless
codes for at least some ranges of interest, and in
\secref{s:numerical} we develop and analyze specific instances of our
codes generated numerically.  In \secref{s:existence}, we show that
within the constraints of our construction rateless codes for any
target ceiling and range can be constructed that are arbitrarily close
to perfect in an appropriate sense.  In \secref{s:desimp} we make some
comments on design and implementation issues, and in \secref{s:sim} we
describe the results of simulations with our constructions.  In
\secref{s:tvc}, we discuss and develop simple extensions of our basic
construction to time-varying channels.  Finally, \secref{s:conc}
provides some concluding remarks.

\section{Background}
\label{s:back}

From a purely information theoretic perspective the problem of
rateless transmission is well understood; see
Shulman~\cite{ShulmanPHD} for a comprehensive treatment.  Indeed, for
channels having one maximizing input distribution, a codebook drawn
independently and identically distributed (i.i.d.) at random from this
distribution will be good with high probability, when truncated to (a
finite number of) different lengths.  Phrased differently, in such
cases random codes are rateless codes.

Constructing good codes that also have computationally efficient
encoders and decoders requires more effort.  A remarkable example of
such codes for \emph{erasure} channels are the recent Raptor codes of
Shokrollahi~\cite{Shokrollahi}, which build on the LT codes of
Luby~\cite{LubyPat,Byers}.  An erasure channel model (for packets) is
most appropriate for rateless coding architectures anchored at the
application layer, where there is little or no access to the physical
layer.

Apart from erasure channels, there is a growing interest in exploiting
rateless codes closer to the physical layer, where AWGN models are
more natural; see, e.g., \cite{Soljanin} and the references therein.
Much less is known about the limits of what is possible in this realm,
which has been the focus of traditional hybrid ARQ research.

One line of work involves extending Raptor code constructions to
binary-input AWGN channels (among others).  In this area,
\cite{Etesami,Palanki} have shown that no degree distribution allows
such codes to approach capacity simultaneously at different signal to
noise ratios (SNRs).  Nevertheless, this does not rule out the
possibility that such codes, when suitably designed, can be near
capacity at multiple SNRs.

A second approach is based on puncturing of low-rate
capacity-approaching binary codes such as turbo and LDPC codes
\cite{Barb,RM,Mantha,Soljanin,SSV,HKM}, or extending a higher-rate
such code, or using a combination of code puncturing and extension
\cite{Narayanan02}.  When iterative decoding is involved, such
approaches lead to performance tradeoffs at different
rates---improving performance at one rate comes at the expense of the
performance at other rates.  While practical codes have been
constructed in this manner \cite{Narayanan02,HKM}, it remains to be
understood how close, in principle, one can come to capacity
simultaneously at multiple SNRs, particularly when not all SNRs are
low.

Finally, for the higher rates typically of interest, which necessitate
higher-order (e.g., 16-QAM and larger) constellations, the modulation
used with such binary codes becomes important.  In turn, such
modulation tends to further complicate the iterative decoding,
imposing additional code design challenges.  Constellation
rearrangement and other techniques have been developed to at least
partially address such challenges
\cite{wengerter,LTE1,LTE2,Doettling}, but as yet do not offer a
complete solution.  Alternatively, suitably designed binary codes can
be, in principle, combined with bit-interleaved coded modulation
(BICM) for such applications; for example, \cite{BarronMilcom}
explores the design of Raptor codes for this purpose, and shows by
example that the gaps to capacity need not be too large, at least
provided the rates are not too high.

From the perspective of the broader body of related work described
above, the present paper represents somewhat of a departure in
approach to the design of rateless codes and hybrid ARQ systems.
However, with this departure come additional complementary insights,
as we will develop.

\section{Channel and System Model}
\label{s:model}

The codes we construct are designed for a complex AWGN channel
\begin{equation}
  \rvv{y}_m = \beta\rvv{x}_m + \rvv{z}_m,\quad m=1,2,\dots,
\label{e:model}
\end{equation}
where $\beta$ is a channel gain,\footnote{More general models for
$\beta$ will be discussed later in the paper.} $\rvv{x}_m$ is a vector
of $N$ input symbols, $\rvv{y}_m$ is the vector of channel output
symbols, and $\rvv{z}_m$ is a noise vector of $N$ i.i.d.\ complex,
circularly-symmetric Gaussian random variables of variance $\sigma^2$,
independent across blocks $m=1,2,\dots$.  The channel input is limited
to average power $P$ per symbol.  In our model, the channel gain
$\beta$ and noise variance $\sigma^2$ are known a priori at the
receiver but not at the transmitter.\footnote{An equivalent model
would be a broadcast channel in which a single encoding of a common
message is being sent to a multiplicity of receivers, each
experiencing a different SNR\@.}

The block length $N$ has no important role in the analysis that
follows.  It is, however, the block length of the base code used in
the rateless construction.  As the base code performance controls the
overall code performance, to approach channel capacity $N$ must be
large.

The encoder transmits a message $w$ by generating a sequence of code
blocks (incremental redundancy blocks) $\rvv{x}_1(w)$, $\rvv{x}_2(w)$,
\dots.  The receiver accumulates sufficiently many received blocks
$\rvv{y}_1$, $\rvv{y}_2$, \dots to recover $w$.  The channel gain $\beta$
may be viewed as a variable parameter in the model; more incremental
redundancy is needed to recover $w$ when $\beta$ is small than when
$\beta$ is large.

An important feature of this model is that the receiver always starts
receiving blocks from index $m=1$.  It does not receive an arbitrary
subsequence of blocks, as might be the case if one were modeling a
broadcast channel that permits ``tuning in'' to an ongoing
transmission.


We now define some basic terminology and notation.  Unless noted
otherwise, all logarithms are base 2, all symbols denote complex
quantities, and all rates are in bits per complex symbol (channel
use), i.e., b/s/Hz.  We use $\cdot\,^\tr$ for transpose and
$\cdot\,^\cgt$ for Hermitian (conjugate transpose) operators.  Vectors
and matrices are denoted using bold face, random variables are denoted
using sans-serif fonts, while sample values use regular (serif) fonts.

We define the \emph{ceiling rate} of the rateless code as the highest
rate $R$ at which the code can operate, i.e., the effective rate if
the message is decoded from the single received block $\rvv{y}_1$;
hence, a message consists of $NR$ information bits.  Associated with
this rate is an SNR \emph{threshold}, which is the minimum SNR
required in the realized channel for decoding to be possible from this
single block.  This SNR threshold can equivalently be expressed in the
form of a channel gain threshold.  Similarly, if the message is
decoded from $m\ge2$ received blocks, the corresponding effective code
rate is $R/m$, and there is a corresponding SNR (and channel gain)
threshold.  Thus, for a rateless encoding consisting of $M$ blocks,
there is a sequence of $M$ associated SNR thresholds.

Finally, as in the introduction, we refer to the code out of which our
rateless construction is built as the \emph{base code}, and the
associated rate of this code as simply the \emph{base code rate}.  At
points in our analysis we assume that a good base code is used in
the code design, i.e., that the base code is capacity-achieving for
the AWGN channel, and thus has the associated properties of such
codes.  This allows us to distinguish losses due to the code
architecture from those due to the choice of base code.

\section{Motivating Example} 
\label{s:example}

To develop initial insights, we construct a simple low-complexity
perfect rateless code that employs two layers of coding to support a
total of two redundancy blocks.  

We begin by noting that for the case of a rateless code with two
redundancy blocks the channel gain $|\beta|$ may be classified into
three intervals based on the number of blocks needed for decoding.
Let $\alpha_1$ and $\alpha_2$ denote the two associated channel gain
thresholds.  When $|\beta|\ge \alpha_1$ decoding requires only one
block.  When $\alpha_1>|\beta|\ge\alpha_2$ decoding requires two
blocks.  When $\alpha_2>|\beta|$ decoding is not possible.  The
interesting cases occur when the gain is as small as possible to
permit decoding.  At these threshold values, for one-block decoding
the decoder sees (aside from an unimportant phase shift)
\begin{equation} 
\rvv{y}_1 = \alpha_1 \rvv{x}_1 + \rvv{z}_1, 
\end{equation}
while
for two-block decoding the decoder sees
\begin{align}
  \rvv{y}_1 &= \alpha_2 \rvv{x}_1 + \rvv{z}_1,\\
  \rvv{y}_2 &= \alpha_2 \rvv{x}_2 + \rvv{z}_2.
\end{align}

In general, given any particular choice of the ceiling rate $R$ for
the code, we would like the resulting SNR thresholds to be a low as
possible.  To determine lower bounds on these thresholds, let 
\begin{equation} 
  \SNR_m = P\alpha_m^2/\sigma^2, \label{e:SNRdef}
\end{equation}
and note that the capacity of the one-block channel is
\begin{equation}
  I_1 = \log (1 + \SNR_1), \label{e:C1}
\end{equation}
while for the two-block channel the capacity is
\begin{equation}
  I_2 = 2\log(1 + \SNR_2) \label{e:C2}
\end{equation}
bits per channel use.  A ``channel use'' in the second case consists
of a pair of transmitted symbols, one from each block.  

In turn, since we deliver the same message to the receiver for both
the one- and two-block cases, the smallest values of $\alpha_1$ and
$\alpha_2$ we can hope to achieve occur when
\begin{equation} 
  I_1=I_2=R.  
  \label{e:perfect2}
\end{equation}
Thus, we say that the code is \emph{perfect} if it is decodable at
these limits.

We next impose that the construction be a \emph{layered} code, and
that the layers be \emph{successively decodable}.

Layering means that we require the transmitted blocks
to be linear combinations of two base codewords $\rvv{c}_1\in
\mathcal{C}_1$ and $\rvv{c}_2\in \mathcal{C}_2$\footnote{In practice,
  the codebooks $\mathcal{C}_1$ and $\mathcal{C}_2$ should not be
  identical, though they can for example be derived from a common base
  codebook via scrambling.  This point is discussed further
  in~\secref{s:desimp}.}:
\begin{align}
  \rvv{x}_1 &= g_{11}\rvv{c}_1 + g_{12}\rvv{c}_2,\\
  \rvv{x}_2 &= g_{21}\rvv{c}_1 + g_{22}\rvv{c}_2.
\end{align}
Base codebook $\mathcal{C}_1$ has rate $R_1$ and base codebook
$\mathcal{C}_2$ has rate $R_2$, where $R_1+R_2=R$, so that total rate
of the two codebooks equals the ceiling rate.  We assume for this
example that both codebooks are capacity-achieving, so that the
codeword components are i.i.d.\ Gaussian.  Furthermore, for
convenience, we scale the codebooks to have unit power, so the power
constraint instead enters through the constraints
\begin{align}
  |g_{11}|^2+|g_{12}|^2 &= P, \\
  |g_{21}|^2+|g_{22}|^2 &= P.
\end{align}
Finally, the successive decoding constraint in our system means that
the layers are decoded one at a time to keep complexity low (on order
of the base code complexity).  Specifically, the decoder first
recovers $\rvv{c}_2$ while treating $\rvv{c}_1$ as additive Gaussian
noise, then recovers $\rvv{c}_1$ using $\rvv{c}_2$ as side
information.

We now show that perfect rateless codes are possible within these
constraints by constructing a matrix $\svv{G}=[g_{ml}]$ so that the
resulting code satisfies \eqref{e:perfect2}.  Finding an admissible
$\svv{G}$ is simply a matter of some algebra: in the one-block case we
need
\begin{align}
  R_1 &= I_{\alpha_1}(\rvs{c}_1;\rvs{y}_1|\rvs{c}_2) \label{e:R1I}\\
  R_2 &= I_{\alpha_1}(\rvs{c}_2;\rvs{y}_1), \label{e:R2I}
\end{align}
and in the two-block case we need
\begin{align}
  R_1 &= I_{\alpha_2}(\rvs{c}_1;\rvs{y}_1,\rvs{y}_2|\rvs{c}_2)\label{e:R1II}\\
  R_2 &= I_{\alpha_2}(\rvs{c}_2;\rvs{y}_1,\rvs{y}_2).\label{e:R2II}
\end{align}
The subscripts $\alpha_1$ and $\alpha_2$ are a reminder that these
mutual information expressions depend on the channel gain, and the
scalar variables denote individual components from the input and
output vectors.

While evaluating \eqref{e:R1I}--\eqref{e:R1II} is straightforward,
calculating the more complicated \eqref{e:R2II}, which corresponds to
decoding $\rvv{c}_2$ in the two-block case, can be circumvented by a
little additional insight.  In particular, while $\rvv{c}_1$ causes
the effective noise in the two blocks to be correlated, observe that a
capacity-achieving code requires $\rvs{x}_1$ and $\rvs{x}_2$ to be
i.i.d.\ Gaussian.  As $\rvs{c}_1$ and $\rvs{c}_2$ are Gaussian,
independent, and equal in power by assumption, this occurs only if the
rows of $\svv{G}$ are orthogonal.  Moreover, the power constraint $P$
ensures that these orthogonal rows have the same norm, which implies
that $\svv{G}$ is a scaled unitary matrix.

The unitary constraint has an immediate important consequence: the
per-layer rates $R_1$ and $R_2$ must be equal, i.e.,
\begin{equation} 
  R_1=R_2=R/2.
  \label{e:equal-R}
\end{equation}
This occurs because the two-block case decomposes into two parallel
orthogonal channels of equal SNR\@.  We see in the next section
that a comparable result holds for any number of layers.

From the definitions of $\SNR_1$ and $I_1$ [cf.\ \eqref{e:SNRdef} and
\eqref{e:C1}], and the equality $I_1=R$ \eqref{e:perfect2}, we find
that
\begin{equation}
  P\alpha_1^2/\sigma^2 = 2^R-1.
  \label{e:fum1}
\end{equation}
Also, from \eqref{e:R1I} and \eqref{e:equal-R}, we find that
\begin{equation}
  |g_{11}|^2\alpha_1^2/\sigma^2 = 2^{R/2}-1.
  \label{e:fum2}
\end{equation}
Combining \eqref{e:fum1} and \eqref{e:fum2} yields
\begin{equation}
  |g_{11}|^2 = P\frac{2^{R/2}-1}{2^R-1} = \frac{P}{2^{R/2}+1}.
\end{equation}
The constraint that $\svv{G}$ be a scaled unitary matrix, together
with the power constraint $P$, implies
\begin{align}
  |g_{12}|^2 & = P - |g_{11}|^2\\
  |g_{21}|^2 & = P - |g_{11}|^2\\
  |g_{22}|^2 & = |g_{11}|^2,
\end{align}
which completely determines the squared modulus of the entries of
$\svv{G}$.

Now, the mutual information expressions \eqref{e:R1I}--\eqref{e:R2II}
are unaffected by applying a common complex phase shift to any row or
column of $\svv{G}$, so without loss of generality we take the first
row and first column of $\svv{G}$ to be real and positive.  For
$\svv{G}$ to be a scaled unitary matrix, $g_{22}$ must then be real
and negative.  We have thus shown that, if a solution to
\eqref{e:R1I}--\eqref{e:R2II} exists, it must have the form
\begin{equation}
\svv{G}=
  \begin{bmatrix}
    g_{11} & g_{12}\\ g_{21} & g_{22}
  \end{bmatrix}
  =
  \sqrt{\frac{P}{2^{R/2}+1}}
  \begin{bmatrix}
    1 & 2^{R/4} \\
    2^{R/4} & -1
  \end{bmatrix}.  \label{e:tbt}
\end{equation}
Conversely, it is straightforward to verify that
\eqref{e:R1I}--\eqref{e:R2II} are satisfied with this selection.  Thus
\eqref{e:tbt} characterizes the (essentially) unique solution
$\svv{G}$.\footnote{Interestingly, the symmetry in \eqref{e:tbt}
implies that the construction remains perfect even if the two
redundancy blocks are received in swapped order.  This is not true of
our other constructions.}

In summary, we have constructed a 2-layer, 2-block perfect rateless
code from linear combinations of codewords drawn from equal-rate
codebooks.  Moreover, decoding can proceed one layer at a time with no
loss in performance, provided the decoder is cognizant of the
correlated noise caused by undecoded layers.  In the sequel we
consider the generalization of our construction to an arbitrary number
of layers and redundancy blocks.

\section{Rateless Codes with Layered Encoding and Successive Decoding}
\label{s:ldr}

The rateless code construction we pursue is as follows
\cite{Erez-perfect}.  First, we choose the range (maximum number $M$
of redundancy blocks), the ceiling rate $R$, the number of layers $L$,
and finally the associated codebooks
$\mathcal{C}_1,\dots,\mathcal{C}_L$.  We will see presently that
the $L$ base codebooks must have equal rate $R/L$ when constructing
perfect rateless codes with $M=L$, and in any case using equal 
rates has the advantage of allowing the codebooks
for each layer to be derived from a single base code.

Given codewords $\rvv{c}_l\in \mathcal{C}_l$, $l=1,\dots,L$, the
redundancy blocks $\rvv{x}_1,\dots,\rvv{x}_M$ take the form
\begin{equation}
  \begin{bmatrix}
    \rvv{x}_1 \\ \vdots \\ \rvv{x}_M
  \end{bmatrix}
  =
  \svv{G}
  \begin{bmatrix}
    \rvv{c}_1\\ \vdots \\ \rvv{c}_L
  \end{bmatrix},
\label{e:rbasic}
\end{equation}
where $\svv{G}$ is an $M\times L$ matrix of complex gains and where
$\rvv{x}_m$ for each $m$ and $\rvv{c}_l$ for each $l$ are row vectors
of length $N$.  The power constraint enters by limiting the rows of
$\svv{G}$ to have squared norm $P$ and by normalizing the codebooks to
have unit power.  With this notation, the elements of the
$m$\/th row of $\svv{G}$ are the weights used in constructing the
$m$\/th redundancy block from the $L$ codewords.\footnote{The $l$\/th
column of $\svv{G}$ also has a useful interpretation.  In particular,
one can interpret the construction as equivalent to a ``virtual''
code-division multiple-access (CDMA) system with $L$ users, each
corresponding to one layer of the rateless code.  With this
interpretation, the signature (spreading) sequence for the $l$\/th
virtual user is the $l$\/th column of $\svv{G}$.}  In the sequel we
use $g_{ml}$ to denote the $(m,l)$\/th entry of $\svv{G}$ and
$\svv{G}_{m,l}$ to denote the upper-left $m\times l$ submatrix of
$\svv{G}$.\footnote{Where necessary, we adopt the convention that
$\svv{G}_{m,0}=\mathbf{0}$.}

\begin{figure}[tbp]
\psfrag{&1}[cc]{time $\longrightarrow$}
\psfrag{&2}[cc]{$\longleftarrow$ layer}
\psfrag{&11}[cc]{$g_{11}\rvv{c}_1$}
\psfrag{&12}[cc]{$g_{21}\rvv{c}_1$}
\psfrag{&13}[cc]{$g_{31}\rvv{c}_1$}
\psfrag{&21}[cc]{$g_{12}\rvv{c}_2$}
\psfrag{&22}[cc]{$g_{22}\rvv{c}_2$}
\psfrag{&23}[cc]{$g_{32}\rvv{c}_2$}
\psfrag{&31}[cc]{$g_{13}\rvv{c}_3$}
\psfrag{&32}[cc]{$g_{23}\rvv{c}_3$}
\psfrag{&33}[cc]{$g_{33}\rvv{c}_3$}
\psfrag{&41}[cc]{$g_{14}\rvv{c}_4$}
\psfrag{&42}[cc]{$g_{24}\rvv{c}_4$}
\psfrag{&43}[cc]{$g_{34}\rvv{c}_4$}
\centerline{\includegraphics[width=3.5in]{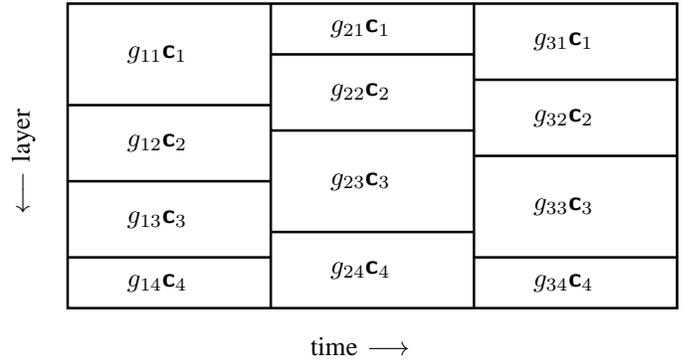}}
\caption{A rateless code construction with 4 layers and 3 blocks of
redundancy.  Each block is a weighted linear combination of the
($N$-element) base codewords $\rvv{c}_1,\rvv{c}_2,\dots,\rvv{c}_4$,
where $g_{ml}$, the $(m,l)$\/th element of $\svv{G}$, denotes the
weight for layer $l$ of block $m$.  In this illustration, the
thickness of a layer is a graphical depiction of the magnitude of its
associated gain (power).
\label{f:construction}}
\end{figure}

An example of this layered rateless code structure is depicted in
Fig.~\ref{f:construction}.  Each redundancy block contains a
repetition of the codewords used in the earlier blocks, but with a
different complex scaling factor.  The code structure may therefore be
viewed as a hybrid of layering and repetition.  Note that, absent
assumptions on the decoder, the order of the layers is not important.

In addition to the layered code structure, there is additional
decoding structure, namely that the layered code be successively
decodable.  Specifically, to recover the message, we first decode
$\rvv{c}_L$, treating $\svv{G} [\rvv{c}_1^\tr \cdots
\rvv{c}_{L-1}^\tr]^\tr$ as (colored) noise, then decode
$\rvv{c}_{L-1}$, treating $\svv{G} [\rvv{c}_1^\tr \cdots
\rvv{c}_{L-2}^\tr]^\tr$ as noise, and so on.  Thus, our aim is to
select $\svv{G}$ so that capacity is achieved for any number
$m=1,\dots,M$ of redundancy blocks subject to the successive decoding
constraint.  Minimum mean-square error (MMSE) combining of the
available redundancy blocks conveniently exploits the repetition
structure in the code when decoding each layer.

Both the layered repetition structure \eqref{e:rbasic} and the
successive decoding constraint impact the degree to which we can
approach a perfect code.  Accordingly, we examine the consequences of
each in turn.

We begin by examining the implications of the layered repetition
structure \eqref{e:rbasic}.  When the number of layers $L$ is at least
as large as the number of redundancy blocks $M$, such layering does
not limit code performance.  But when $L<M$, it does.  In particular,
whenever the number $m$ of redundancy blocks required by the realized
channel exceeds $L$, there is necessarily a gap between the code
performance and capacity.  To see this, observe that \eqref{e:rbasic}
with \eqref{e:model}, restricted to the first $m$ blocks, defines a
linear $L$-input $m$-output AWGN channel, the capacity of which is at
most
\begin{equation}
  I'_m =
  \begin{cases}
    m\log\left(1+\frac{|\beta|^2 P}{\sigma^2}\right) & \text{for $m\le L$,}\\
    L\log\left(1+\frac{m}{L} \frac{|\beta|^2 P}{\sigma^2}\right) &
    \text{for $m>L$.} 
  \end{cases} 
\label{e:Rm}
\end{equation}
Only for $m\le L$ does this match the capacity of a general $m$-block
AWGN channel, viz.,
\begin{equation}
  I_m = m\log\left(1+\frac{|\beta|^2 P}{\sigma^2} \right).
\label{e:Cm}
\end{equation}
Ultimately, for $m>L$ the problem is that an $L$-fold linear
combination cannot fill all degrees of freedom afforded by the
$m$-block channel.

An additional penalty occurs when we combine the layered repetition
structure with the requirement that the code be rateless.
Specifically, for $M>L$, there is no choice of gain matrix $\svv{G}$
that permits \eqref{e:Rm} to be met with equality
\emph{simultaneously} for all $m=1,\dots,M$.  A necessary and
sufficient condition for equality is that the rows of $\svv{G}_{m,L}$
be orthogonal for $m\le L$ and the columns of $\svv{G}_{m,L}$ be
orthogonal for $m>L$.  This follows because reaching \eqref{e:Rm} for
$m\le L$ requires that the linear combination of $L$ codebooks create
an i.i.d.\ Gaussian sequence.  In contrast, reaching \eqref{e:Rm} for
$m> L$ requires that the linear combination inject the $L$ codebooks
into orthogonal subspaces, so that a fraction $L/m$ of the available
degrees of freedom are occupied by i.i.d.\ Gaussians (the rest being
empty).

Unfortunately, the columns of $\svv{G}_{m,L}$ cannot be orthogonal
simultaneously for all $m>L$; orthogonal $m$-dimensional vectors
(with nonzero entries) cannot remain
orthogonal when truncated to their first $m-1$ dimensions.
Thus \eqref{e:Rm} determines only a lower bound on the
loss due to the layering structure \eqref{e:rbasic}.  Fortunately, the
additional loss encountered in practice turns out to be quite small,
as we demonstrate numerically as part of the next section.

When $M=L$, the orthogonality requirement forces $\svv{G}$ to be a
scaled unitary matrix.  Upon receiving the final redundancy block $m=M$,
the problem decomposes into $L$ parallel channels
with equal SNR, which in turn implies that the rate of each
layer must equal $R/L$.

A lower bound on loss incurred by the use of insufficiently many
layers is readily obtained by comparing \eqref{e:Rm} and \eqref{e:Cm}.
Given a choice of ceiling rate $R$ for the rateless
code, \eqref{e:Rm} implies that for rateless codes constructed using
linear combinations of $L$ base codes, the smallest channel gain
$\alpha'_m$ for which it's possible to decode with $m$ blocks is
\begin{equation}
  {\alpha^{\prime 2}_m} =
  \begin{cases}
    \left(2^{R/m} - 1\right) \frac{\sigma^2}{P} & \text{for $m\le L$,} \\
    \left(2^{R/L} - 1\right) \frac{L}{m} \frac{\sigma^2}{P} & \text{for $m>L$.}
  \end{cases} 
\label{e:modalpha}
\end{equation}
By comparison, \eqref{e:Cm} implies that without the layering
constraint the corresponding channel gain thresholds $\alpha_m$ are
\begin{equation}
  \alpha_m^2 = \left(2^{R/m} - 1\right)\frac{\sigma^2}{P}.
\label{e:alpha}
\end{equation}

The resulting performance loss $\alpha'_m/\alpha_m$ caused by the
layered structure as calculated from \eqref{e:modalpha} and
\eqref{e:alpha} is shown in dB in Table~\ref{t:llosses} for a target
ceiling rate of $R=5$ bits/symbol.  For example, if an application
requires $M=10$ redundancy blocks, a 3-layer code has a loss of less
than 2 dB at $m=10$, while a 5-layer code has a loss of less than 0.82
dB at $m=10$.
\begin{table}
  \caption{Losses $\alpha'_m/\alpha_m$ in dB due to layered
  structure imposed on a rateless code of ceiling rate $R=5$ 
  b/s/Hz, as a function of the number of layers $L$ and redundancy
  blocks $m$. \label{t:llosses}} 
 \begin{center}
  \begin{tabular}{@{\extracolsep{-1pt}}cccccccccccc}
  & \multicolumn{9}{c}{Redundancy blocks $m$}\\
    & 2 & 3 & 4 & 5 & 6 & 7& 8& 9 & 10\\
$L=1$ &  5.22 &  6.77 &  7.50 &  7.92 &  8.20 &  8.40 &  8.54 &  8.65 &  8.74 \\
$L=2$ &  0.00 &  1.55 &  2.28 &  2.70 &  2.98 &  3.17 &  3.32 &  3.43 &  3.52 \\
$L=3$ &  0.00 &  0.00 &  0.73 &  1.16 &  1.43 &  1.63 &  1.77 &  1.88 &  1.97 \\
$L=4$ &  0.00 &  0.00 &  0.00 &  0.42 &  0.70 &  0.90 &  1.04 &  1.15 &  1.24 \\
$L=5$ &  0.00 &  0.00 &  0.00 &  0.00 &  0.28 &  0.47 &  0.62 &  0.73 &  0.82 \\
$L=6$ &  0.00 &  0.00 &  0.00 &  0.00 &  0.00 &  0.20 &  0.34 &  0.45 &  0.54 \\
$L=7$ &  0.00 &  0.00 &  0.00 &  0.00 &  0.00 &  0.00 &  0.14 &  0.26 &  0.35 \\
$L=8$ &  0.00 &  0.00 &  0.00 &  0.00 &  0.00 &  0.00 &  0.00 &  0.11 &  0.20 \\
$L=9$ &  0.00 &  0.00 &  0.00 &  0.00 &  0.00 &  0.00 &  0.00 &  0.00 &  0.09 \\
  \end{tabular}
 \end{center}
\end{table}

As Table~\ref{t:llosses} reflects---and as can be readily
verified analytically---for a fixed number of layers $L$
and a fixed base code rate
$R/L$, the performance loss $\alpha'_m/\alpha_m$ attributable to
the imposition of layered encoding grows monotonically with the number
of blocks $m$, approaching the limit
\begin{equation} 
\frac{\alpha_\infty^{\prime 2}}{\alpha_\infty^2} = \frac{2^{R/L}-1}{(R/L)\ln
  2}.
\end{equation}
Thus, in applications where the number of incremental redundancy
blocks is very large, it's advantageous to keep the base code rate
small.  For example, with a base code rate of 1/2 bit per complex
symbol (implemented, for example, using a rate-1/4 binary code) the
loss due to layering is at most 0.78 dB, while with a base code rate
of 1 bit per complex symbol the loss is at most 1.6 dB\@.

We now determine the additional impact the successive decoding
requirement has on our ability to approach capacity, and more
generally what constraints it imposes on $\svv{G}$.  We continue to
incorporate the power constraint by taking the rate-$R/L$ codebooks
$\mathcal{C}_1,\dots,\mathcal{C}_L$ to have unit power and the rows of
$\svv{G}$ to have squared norm $P$.  Since our aim is to employ
codebooks designed for (non-fading) Gaussian channels, we make the
further assumption that the codebooks have constant power, i.e., that
they satisfy the per-symbol energy constraint
$E\bigl[|c_{l,n}(\rvs{w})|^2\bigr]\le 1$ for all layers $l$ and time indices
$n=1,\dots,N$, where the expectation is taken over equiprobable
messages $\rvs{w}\in\{1,\dots,2^{NR/L}\}$.  Additional constraints on
$\svv{G}$ now follow from the requirement that the mutual
information accumulated through any block $m$ at each layer $l$ be
large enough to permit successive decoding.

Concretely, suppose we have received blocks $1,\dots,m$.  Let the
optimal threshold channel gain $\alpha_m$ be defined as in
\eqref{e:alpha}.
Suppose further that layers $l+1,\dots,L$ have been successfully
decoded, and define
\begin{equation} 
\begin{bmatrix}\rvv{v}_{1} \\ \vdots \\ \rvv{v}_{m} \end{bmatrix}
 = \beta \svv{G}_{m,l}
   \begin{bmatrix}\rvv{c}_1 \\ \vdots \\ \rvv{c}_l\end{bmatrix} +
  \begin{bmatrix}\rvv{z}_1 \\ \vdots \\ \rvv{z}_m\end{bmatrix}
\label{e:vl-def}
\end{equation}
as the received vectors without the contribution from layers
$l+1,\dots,L$.

Then, following standard arguments, with independent equiprobable
messages for each layer, the probability of decoding error for layers
$1,\dots,l$ can made vanishingly small with increasing block length only if
the mutual information between input and output is at least as large
as the combined rate $lR/L$ of the codes $\mathcal{C}_1,\dots,\mathcal{C}_l$.
That is, when $\beta$ equals the optimal threshold gain $\alpha_m$,
successive decoding requires
\begin{align}
  lR/L
  &\le (1/N)I(\rvv{c}_1,\dots,\rvv{c}_l; \rvv{y}_1,\dots,\rvv{y}_m \mid \rvv{c}_{l+1}^{L}) \label{e:foo1}\\
  &=  (1/N)
  I(\rvv{c}_1,\dots,\rvv{c}_l;\rvv{v}_{1},\dots,\rvv{v}_{m}) \label{e:foo2}\\
  &= (1/N)(H(\rvv{v}_1,\dots,\rvv{v}_m) - H(\rvv{v}_1,\dots,\rvv{v}_m|\rvv{c}_1,\dots,\rvv{c}_l))\\
  &\le \log\det(\sigma^2\svv{I} + \alpha_m^2 \svv{G}_{m,l}\svv{G}_{m,l}^\cgt) - \log\det (\sigma^2\svv{I}) \label{e:foo3}\\
  &= \log\det(\svv{I}+(\alpha_m^2/\sigma^2) \svv{G}_{m,l}\svv{G}_{m,l}^\cgt), \label{e:layerinf}
\end{align}
where $\svv{I}$ is an appropriately sized ($m\times m$) identity
matrix.  The inequality~\eqref{e:foo3} relies on the assumption
that the codebooks have constant power, and it holds with equality if
the components of $\svv{G}_{m,l}[\rvv{c}_1^T,\dots,\rvv{c}_l^T]^T$ 
are jointly Gaussian, which by Cramer's theorem requires the components
of $\rvv{c}_1,\dots,\rvv{c}_l$ to be jointly Gaussian.


Our ability to choose $\svv{G}$ to either exactly or approximately
satisfy \eqref{e:layerinf} for all $l=1,\dots,L$ and each
$m=1,\dots,M$ determines the degree to which we can approach capacity.
It is straightforward to see that there is no slack in the problem;
\eqref{e:layerinf} can be satisfied simultaneously for all $l$ and $m$
only if the inequalities are all met with equality.  Beyond this
observation, however, the conditions under which \eqref{e:layerinf}
may be satisfied are not obvious.

Characterizing the set of solutions for $\svv{G}$ when $L=M=2$ was
done in \secref{s:example} (see \eqref{e:tbt}).  Characterizing
the set of solutions when $L=M=3$ requires more work.  It is shown in
Appendix~\ref{a:rateless33} that, when it exists, a solution $\svv{G}$
must have the form
\begin{multline}
  \svv{G} = \sqrt{x-1} \cdot{} \\
  \begin{bmatrix}
    \sqrt{x+1} & \sqrt{x^2(x+1)} & \sqrt{x^4(x+1)} \\
    \sqrt{x^3(x+1)} & e^{j\theta_1}\sqrt{x^5+1} & e^{j\theta_2}\sqrt{x(x+1)} \\
    \sqrt{x^2(x^3+1)} & e^{j\theta_3}\sqrt{x(x^3+1)} & e^{j\theta_4}\sqrt{x^3+1}
  \end{bmatrix} \label{e:ratelessm33}
\end{multline}
where $x=2^{R/6}$ and where $e^{j\theta_i}$, $i=1,\dots,4$ are complex
phasors.  The desired phasors---or a proof of nonexistence---may be
determined from the requirement that $\svv{G}$ be a scaled unitary
matrix.  Using this observation, it is shown in
Appendix~\ref{a:rateless33} that a solution $\svv{G}$ exists and is
unique (up to complex conjugate) for all $R\le 3(\log (7+3\sqrt{5})
-1)\approx 8.33$ bits per complex symbol, but no choice of phasors
results in a unitary $\svv{G}$ for larger values of $R$.

For example, using \eqref{e:ratelessm33} with $R=6$ bits/symbol we
find that:
\begin{equation*}
  P = 63,\quad \alpha_1 = 1,\quad \alpha_2 = \sqrt{1/9}, \quad \alpha_3 = \sqrt{1/21}
\end{equation*}
\begin{displaymath}
  \svv{G} =
  \begin{bmatrix}
    \sqrt{3} &  \sqrt{12}  & \sqrt{48} \\
    \sqrt{24} & \sqrt{33} e^{j \theta_1}  & \sqrt{6} e^{j \theta_2}  \\
    \sqrt{36} & \sqrt{18} e^{j\theta_3} & \sqrt{9} e^{j\theta_4}
  \end{bmatrix}
\end{displaymath}
where
\begin{align*}
  \theta_1 &= \arccos\frac{-5}{2\sqrt{22}},&
  \theta_2 &= 2\pi - \arctan{3\sqrt{7}}, \\
  \theta_3 &= -\arctan{\sqrt{7}},&
  \theta_4 &= \pi - \arctan{\sqrt{7}/3}.
\end{align*}

For $M>3$ the algebra becomes daunting, though we conjecture that
exact solutions and hence perfect rateless codes exist for all $L=M$,
for at least some nontrivial values of $R$.\footnote{In recent
calculations following the above approach, Ayal Hitron at Tel Aviv
University has determined that exact solutions exist in the $M=L=4$
case for rates in the range $R\le 10.549757$.}

For $L<M$ perfect constructions cannot exist.  As developed earlier in
this section, even if we replace the optimum threshold channel gains
$\alpha_m$ defined via \eqref{e:alpha} with suboptimal gains $\alpha'_m$
of \eqref{e:modalpha}
determined by the layering bound \eqref{e:Rm},
it is still not possible to satisfy \eqref{e:layerinf}.  However, one
can come close.  While
the associated analysis is nontrivial, such behavior is easily
demonstrated numerically, which we show as part of the next section.

\section{Numerical Examples}
\label{s:numerical}

In this section, we consider numerical constructions both for the case
$L=M$ and for the case $L<M$.  Specifically, we have experimented
with numerical optimization methods to satisfy \eqref{e:layerinf} for
up to $M=10$ redundancy blocks, using the threshold channel gains
$\alpha'_m$ defined via \eqref{e:modalpha} in place of those defined
via \eqref{e:alpha} as appropriate when the number of blocks $M$
exceeds the number of layers $L$.

For the case $L=M$, for each of $M=2,3,\dots,10$, we found
constructions with $R/L=2$ bits/symbol that come within 0.1\% of
satisfying \eqref{e:layerinf} subject to \eqref{e:alpha}, and often
the solutions come within 0.01\%.  This provides powerful evidence
that perfect rateless codes exist for a wide range of parameter
choices.

For the case $L<M$, despite the fact that there do not exist perfect
codes, in most cases of interest one can come remarkably close to
satisfying \eqref{e:layerinf} subject to \eqref{e:modalpha}.
Evidently mutual information for Gaussian channels is quite
insensitive to modest deviations of the noise covariance away from a
scaled identity matrix.

As an example, Table~\ref{t:shortfall} shows the rate shortfall in
meeting the mutual information constraints \eqref{e:layerinf} for an
$L=3$ layer code with $M=10$ redundancy blocks, and a target ceiling
rate $R=5$.  The associated complex gain matrix is
\begin{equation*}
\svv{G} = \left[
\begin{array}{lll}
    1.4747 & 2.6277 & 4.6819 \\
    3.5075 & 3.7794\,e^{j2.0510} & 2.1009\,e^{-j1.9486} \\
    4.0648 & 3.1298\,e^{-j0.9531} & 2.1637\,e^{j2.5732} \\
    3.2146 & 3.1322\,e^{j3.0765} & 3.2949\,e^{j0.9132} \\
    3.2146 & 3.3328\,e^{-j1.6547} & 3.0918\,e^{-j1.4248} \\
    3.2146 & 3.1049\,e^{j0.9409} & 3.3206\,e^{j2.8982} \\
    3.2146 & 3.3248\,e^{j1.2506} & 3.1004\,e^{-j0.2027} \\
    3.2146 & 3.0980\,e^{-j1.4196} & 3.3270\,e^{j1.9403} \\
    3.2146 & 3.2880\,e^{-j2.9449} & 3.1394\,e^{-j1.9243} \\
    3.2146 & 3.1795\,e^{j0.7839} & 3.2492\,e^{j0.3413}
\end{array}
\right].
\end{equation*}
The worst case loss is less than 1.5\%; this example is typical in
its efficiency.

\begin{table}
  \caption{Percent shortfall in rate for a numerically-optimized
  rateless code with $M=10$ blocks, $L=3$ layers, and a ceiling rate
  of $R=5$ b/s/Hz. \label{t:shortfall}}  
\begin{center}
\begin{tabular}{@{\extracolsep{-3pt}}cccccccccccc}
  & \multicolumn{9}{c}{Redundancy blocks $m$}\\
   & 1 & 2 & 3 & 4 & 5 & 6 & 7& 8& 9 & 10\\
$l=1$ & 0.00 & 0.00 & 0.00 & 0.00 & 0.00 & 0.00 & 0.00 & 0.00 & 0.00 & 0.00 \\
$l=2$ & 0.00 & 0.28 & 1.23 & 1.46 & 1.39 & 0.44 & 0.59 & 0.48 & 0.16 & 0.23 \\
$l=3$ & 0.00 & 0.29 & 1.23 & 1.48 & 1.40 & 0.43 & 0.54 & 0.51 & 0.15 & 0.23
\end{tabular}
\end{center}
\end{table}

The total loss of the designed code relative to a perfect rateless
code is, of course, the sum of the successive decoding and layered
encoding constraint losses.  Hence, the losses in
Tables~\ref{t:llosses} and \ref{t:shortfall} are cumulative.  As a
practical matter, however, when $L<M$, the layered encoding constraint
loss dwarfs that due to the successive decoding constraint: the
overall performance loss arises almost entirely from the code's
inability to occupy all available degrees of freedom in the channel.
Thus, this overall loss can be estimated quite closely by comparing
\eqref{e:Cm} and \eqref{e:Rm}.  Indeed this is reflected in our example,
where the loss of Table~\ref{t:llosses} dominates over that of
Table~\ref{t:shortfall}.

\section{Existence of Near-Perfect Rateless Codes}
\label{s:existence}

While the closed-form construction of \emph{perfect} rateless codes
subject to layered encoding and successive decoding becomes more
challenging with increasing code range $M$, the construction of codes
that are at least nearly perfect is comparatively straightforward.  In
the preceding section, we demonstrated this numerically.  In this
section, we prove this analytically.  In particular, we construct
rateless codes that are arbitrarily close to perfect in an appropriate
sense, provided enough layers are used.  We term these near-perfect
rateless codes. The code construction we present is applicable to
arbitrarily large $M$ and also allows for simpler decoding than that
required in the preceding development.

The near-perfect codes we develop in this section \cite{Erez} are
closely related to those in \secref{s:ldr}.  However, there are a few
differences.  We retain the layered construction, but instead of using
a single complex weight for the codeword at each layer (and block), we
use a single weight magnitude for each codeword and vary the phase of
the weight from symbol to symbol within the codeword in each layer
(and block).  Moreover, in our analysis, the phases are chosen
randomly, corresponding to evaluating an ensemble of codes.  The
realizations of these random phases are known to and exploited by the
associated decoders.  As with the usual random coding development, we
establish the existence of good codes in the ensemble by showing that
the average performance is good.

These modifications, and in particular the additional degrees of
freedom in the code design, simplify the analysis---at the expense of
some slightly more cumbersome notation.  Additionally, because of
these differences, the particular gain matrices in this section cannot
be easily compared with those of \secref{s:ldr}, but we do not require
such comparisons.

\subsection{Encoding}
\label{s:enc}

As discussed above, in our approach to perfect constructions in
\secref{s:ldr}, we made each redundancy block a linear combination of
the base codewords, where the weights are the corresponding row of the
combining matrix $\svv{G}$, as \eqref{e:rbasic} indicates.  Each
individual symbol of a particular redundancy block is,
therefore, a linear combination of the corresponding symbols in the
respective base codewords, with the combining matrix being the same
for all such symbols.

Since for the codes of this section we allow the combining matrix to
vary from symbol to symbol in the construction of each redundancy
block, we augment our notation.  In particular, using $\rvs{c}_l(n)$
and $\rvs{x}_m(n)$ to denote the $n$\/th elements of codeword
$\rvv{c}_l$ and redundancy block $\rvv{x}_m$, respectively, we have
[cf.\ \eqref{e:rbasic}]
\begin{equation}
  \begin{bmatrix}
    \rvs{x}_1(n) \\ \vdots \\ \rvs{x}_M(n)
  \end{bmatrix}
  =
  \rvv{G}(n)
  \begin{bmatrix}
    \rvs{c}_1(n)\\ \vdots \\ \rvs{c}_L(n)
  \end{bmatrix},\quad n=1,2,\dots,N.
\label{e:rbasic2}
\end{equation}
The value of $M$ plays no role in our development and may be
taken arbitrarily large.  Moreover, as before, the power constraint
enters by limiting the rows of $\rvv{G}(n)$ to have a squared norm $P$
and by normalizing the codebooks to have unit power.

It suffices to restrict our attention to $\rvv{G}(n)$ of the
form
\begin{equation}
\rvv{G}(n)= \svv{P} \odot \rvv{D}(n), 
\label{e:PD_form}
\end{equation}
where $\svv{P}$ is an $M\times L$ (deterministic) power allocation
matrix with entries $\sqrt{p_{m,l}}$ that do not vary within a block,
\begin{equation}
  \svv{P} = 
  \begin{bmatrix}
    \sqrt{p_{1,1}} & \ldots & \sqrt{p_{1,L}}\\
    \vdots & \ddots & \vdots\\
    \sqrt{p_{M,1}} & \ldots & \sqrt{p_{M,L}}
  \end{bmatrix},
\label{e:P}
\end{equation}
and $\rvv{D}(n)$ is a (random) phase-only ``dither'' matrix of the
form
\begin{align}
  \rvv{D}(n) &=
   \begin{bmatrix}
    \rvs{d}_{1,1}(n) & \cdots & \rvs{d}_{1,L}(n)\\
    \vdots & \ddots & \vdots\\
    \rvs{d}_{M,1}(n) & \cdots & \rvs{d}_{M,L}(n)
  \end{bmatrix},
\label{e:D}
\end{align}
with $\odot$ denoting elementwise multiplication.  In our analysis,
the $\rvs{d}_{ij}(n)$ are all i.i.d.\ in $i$, $j$, and $n$, and are
independent of all other random variables, including noises,
messages, and codebooks.  As we shall see below, the role of the
dither is to decorrelate pairs of random variables, hence it suffices
for $\rvs{d}_{ij}(n)$ to take values $+1$ and $-1$ with equal
probability.

\subsection{Decoding}
\label{s:dec}

To obtain a near-perfect rateless code, it is sufficient to employ a
successive cancellation decoder with simple maximal ratio combining (MRC) of
the redundancy blocks.  While, in principle, an MMSE-based successive
cancellation decoder enables higher performance, as we will see, an
MRC-based one is sufficient for our purposes, and simplifies the
analysis.  Indeed, although the encoding we choose creates a per-layer
channel that is time-varying, the MRC-based successive cancellation
decoder effectively transforms the channel back into a time-invariant
one, for which any of the traditional low-complexity
capacity-approaching codes for the AWGN channel are suitable as a base
code in the design.\footnote{More generally, the MRC-based decoder is
particularly attractive for practical implementation.  Indeed, as each
redundancy block arrives a sufficient statistic for decoding can be
accumulated without the need to retain earlier blocks in buffers.
The computational cost of decoding thus grows linearly with block
length while the memory requirements do not grow at all.  This is much
less complex than the MMSE decoder discussed in the development of the
codes of \secref{s:ldr}.}

The decoder operation is as follows, assuming the SNR is such that
decoding is possible from $m$ redundancy blocks.  To decode the $L$th
(top) layer, the dithering is first removed from the received waveform
by multiplying by the conjugate dither sequence for that layer.  Then,
the $m$ blocks are combined into a single block via the appropriate
MRC for that layer.  The message in this $L$th layer is then decoded,
treating the undecoded layers as noise, and its contribution
subtracted from the received waveform.  The $(L-1)$\/st layer is now
the top layer, and the process is repeated, until all layers have been
decoded.  Note that the use of MRC in decoding is equivalent to
treating the undecoded layers as white (rather than structured) noise,
which is the natural approach when the dither sequence structure in
those undecoded (lower) layers is ignored in decoding the current
layer of interest.

We now introduce notation that allows the operation of the decoder to
be expressed more precisely.  We then determine the effective SNR seen
by the decoder at each layer of each redundancy block.

Since $\rvv{G}(n)$ is drawn i.i.d., the overall channel is i.i.d., and
thus we may express the channel model in terms of an arbitrary
individual element in the block.  Specifically, our received waveform
can be expressed as [cf.\ \eqref{e:model} and \eqref{e:rbasic}]
\begin{equation}
\rvv{y} = 
\begin{bmatrix}
 \rvs{y}_1\\ \vdots\\ \rvs{y}_M
  \end{bmatrix}
   = \beta \rvv{G} 
  \begin{bmatrix}
    \rvs{c}_1\\ \vdots\\ \rvs{c}_L
  \end{bmatrix}
  +
  \begin{bmatrix}
    \rvs{z}_1\\ \vdots\\ \rvs{z}_M
  \end{bmatrix},
\label{e:rx}
\end{equation}
where $\rvv{G}= \svv{P}\odot \rvv{D}$, with $\rvv{G}$ denoting the
arbitrary element in the sequence $\rvv{G}(n)$, and where
$\rvs{y}_m$ is the corresponding received symbol from redundancy block
$m$ (and similarly for $\rvs{c}_l$, $\rvs{z}_m$, $\rvv{D}$).

If layers $l+1,l+2,\dots,L$ have been successively decoded from $m$
redundancy blocks, and their effects subtracted from the received
waveform, the residual waveform is denoted by
\begin{equation}
  \rvv{v}_{m,l} =
    \beta \rvv{G}_{m,l}\begin{bmatrix}\rvs{c}_1\\ \vdots \\ \rvs{c}_l
    \end{bmatrix}
    + \begin{bmatrix} \rvs{z}_1 \\ \vdots \\ \rvs{z}_m \end{bmatrix},
\end{equation}
where we continue to let $\rvv{G}_{m,l}$ denote the $m\times l$
upper-left submatrix of $\rvv{G}$, and likewise for $\rvv{D}_{m,l}$
and $\svv{P}_{m,l}$.  As additional notation, we let $\rvv{g}_{m,l}$
denote the $m$-vector formed from the upper $m$ rows of the $l$\/th
column of $\rvv{G}$, whence
\begin{equation} 
\rvv{G}_{m,l} = \begin{bmatrix} \rvv{g}_{m,1} & \rvv{g}_{m,2} & \cdots &
  \rvv{g}_{m,l} \end{bmatrix},
\end{equation}
and likewise for $\rvv{d}_{m,l}$ and $\svv{p}_{m,l}$.  

With such notation, the decoding can be expressed as follows.
Starting with $\rvv{v}_{m,L}=\rvv{y}$, decoding proceeds.  After layers
$l+1$ and higher have been decoded and removed, we decode from
$\rvv{v}_{m,l}$.  Writing
\begin{equation}
  \rvv{v}_{m,l} = \beta (\rvv{d}_{m,l}\odot\svv{p}_{m,l})\rvs{c}_l + \rvv{v}_{m,l-1},
\end{equation}
the operation of removing the dither can be expressed as
\begin{equation}
  \rvv{d}_{m,l}^* \odot \rvv{v}_{m,l} = \beta \svv{p}_{m,l} \rvs{c}_l + \rvv{v}'_{m,l-1}
  \label{e:vpp-implicit}
\end{equation}
where 
\begin{equation} 
  \rvv{v}'_{m,l-1}=\rvv{d}_{m,l}^*\odot\rvv{v}_{m,l-1}.
  \label{e:vpp-def}
\end{equation}
The MRC decoder treats the dither in the same manner as noise, i.e.,
as a random process with known statistics but unknown realization.
Because the entries of the dither matrix are chosen to be i.i.d.\
random phases independent of the messages, the entries of
$\rvv{D}_{m,l}$ and $\begin{bmatrix} \rvs{c}_1 & \cdots &
  \rvs{c}_{l-1} \end{bmatrix}$ are jointly and individually
uncorrelated, and the effective noise $\rvv{v}'_{m,l-1}$ seen by the MRC
decoder has diagonal covariance $\svv{K}_{\rvv{v}'_{m,l-1}} =
E[\rvv{v'}_{m,l-1}\rvv{v'}_{m,l-1}^\cgt]$.

The effective SNR at which this $l$\/th layer is decoded from $m$
blocks via MRC is thus
\begin{equation}
\SNR_\mathrm{MRC} = \sum_{m'=1}^{m} \SNR_{m',l}(\beta),
\label{e:post-mrc-snr}
\end{equation}
where
\begin{equation} 
  \SNR_{m',l}(\beta) = \frac{|\beta|^2
  p_{m',l}}{|\beta|^2(p_{m',1}+\dots+p_{m',l-1}) + \sigma^2}.
\label{e:SNRml}
\end{equation}
Note that we have made explicit the dependency of these per-layer per-block
SNRs on $\beta$.

\subsection{Efficiency}
\label{s:eff}

The use of random dither at the encoder and MRC at the decoder both
cause some loss in performance relative to the perfect rateless codes
presented earlier.  In this section we show that these losses can be
made small.

When a coding scheme is not perfect, its \emph{efficiency} quantifies
how close the scheme is to perfect.  There are ultimately several ways
one could measure efficiency that are potentially useful for
engineering design.  Among these, we choose the following efficiency
notion:
\begin{enumerate}
\item We find the ideal thresholds $\{\alpha_m\}$ for a perfect code
  of rate $R$.
\item We determine the highest rate $R'$ such that an imperfect code
  designed at rate $R'$ is decodable with $m$ redundancy blocks when
  the channel gain is $\alpha_m$, for all $m=1,2,\dots$.
\item We measure efficiency $\eta$ by the ratio $R'/R$, which is
  always less than unity.
\end{enumerate}
With this notion of efficiency, we further define a coding scheme as
near-perfect if the efficiency so-defined approaches unity when
sufficiently many layers $L$ are employed.

The efficiency of our scheme ultimately depends on the choice of our
power allocation matrix \eqref{e:P}.  We now show the main result of
this section: provided there exists a power allocation matrix such
that for each $l$ and $m$
\begin{equation}
  \frac{R}{L} = \sum_{m'=1}^m \log(1+\SNR_{m',l}(\alpha_m)),
\label{e:pa-cond}
\end{equation}
with $\SNR_{m,l}(\cdot)$ as defined in \eqref{e:SNRml}, a near-perfect
rateless coding scheme results.  We prove the existence of such a
power allocation---and develop an interpretation of
\eqref{e:pa-cond}---in Appendix~\ref{a:power}, and thus focus on our
main result in the sequel.

We establish our main result by finding a lower bound on the average
mutual information between the input and output of the channel.  Upon
receiving $m$ blocks with channel gain $\alpha_m$, and assuming layers
$l+1,\dots,L$ are successfully decoded, let $I'_{m,l}$ be the mutual
information between the input to the $l$th layer and the channel
output.  Then
\begin{align}
  I'_{l,m} 
  &= I(\rvs{c}_l; \rvv{v}_{m,l} \mid  \rvv{d}_{m,l})  \\
  &= I(\rvs{c}_l; \alpha_m \svv{p}_{m,l} \rvs{c}_l +
  \rvv{v}'_{m,l-1}\mid \rvv{d}_{m,l}),\label{e:bnd2}\\
  &\ge I(\rvs{c}_l; \alpha_m \svv{p}_{m,l} \rvs{c}_l + \rvv{v}'_{m,l}), \label{e:bnd3}\\
  &\ge I(\rvs{c}_l; \alpha_m \svv{p}_{m,l} \rvs{c}_l + \rvv{v}''_{m,l}), \label{e:bnd4}\\
  &= \log\left(1+  \SNR_\mathrm{MRC} \right) \label{e:eval}
\end{align}
where \eqref{e:bnd2} follows from
\eqref{e:vpp-implicit}--\eqref{e:vpp-def}, \eqref{e:bnd3} follows from
the independence of $\rvs{c}_l$ and $\rvv{d}_{m,l}$, and
\eqref{e:bnd4} obtains by replacing $\rvv{v}'_{m,l-1}$ with a Gaussian
random vector $\rvv{v}''_{m,l-1}$ of covariance
$\svv{K}_{\rvv{v}'_{m,l-1}}$.  Lastly, to obtain \eqref{e:eval} we have
used \eqref{e:post-mrc-snr} for the post-MRC SNR\@.

Now, if the assumption \eqref{e:pa-cond} is satisfied, then the
right-hand side of \eqref{e:eval} is further bounded for all $m$
by
\begin{equation} 
  I'_{m,l} \ge \log\left(1+\ln 2\frac{R}{L}\right), \label{e:sand}
\end{equation}
where we have applied the inequality $\ln(1+u)\le u$ (valid for $u>0$)
to \eqref{e:pa-cond} to conclude that $(\ln 2)R/L \le \sum_{m'=1}^m
\SNR_{m',l}(\alpha_m)$.  Note that the lower bound
\eqref{e:sand} may be quite loose; for example, $I'_{m,l}=R/L$ when
$m=1$.

Thus, if we design each layer of the code for a base code rate of
\begin{equation}
  \frac{R''}{L} = \log\left(1+\ln2\frac{R}{L}\right),
\label{e:conserve}
\end{equation}
\eqref{e:sand} ensures decodability after $m$ blocks are received
when the channel gain is $\alpha_m$, for $m=1,2,\dots$.

Finally, rewriting \eqref{e:conserve} as
\begin{equation}
  \frac{R}{L} = \frac{2^{R''/L}-1}{\ln2},
\end{equation}
the efficiency $\eta$ of the conservatively-designed layered
repetition code is bounded by
\begin{equation}
  \eta = \frac{R''}{R} =
   \frac{(\ln2)R''/L}{2^{R''/L}-1} \ge 1-\frac{\ln2}{2}\frac{R''}{L},
\label{e:eff-bnds}
\end{equation}
which approaches unity as $L\to\infty$ as claimed.

\begin{figure}[tbp]
  \centerline{\includegraphics[width=3.5in]{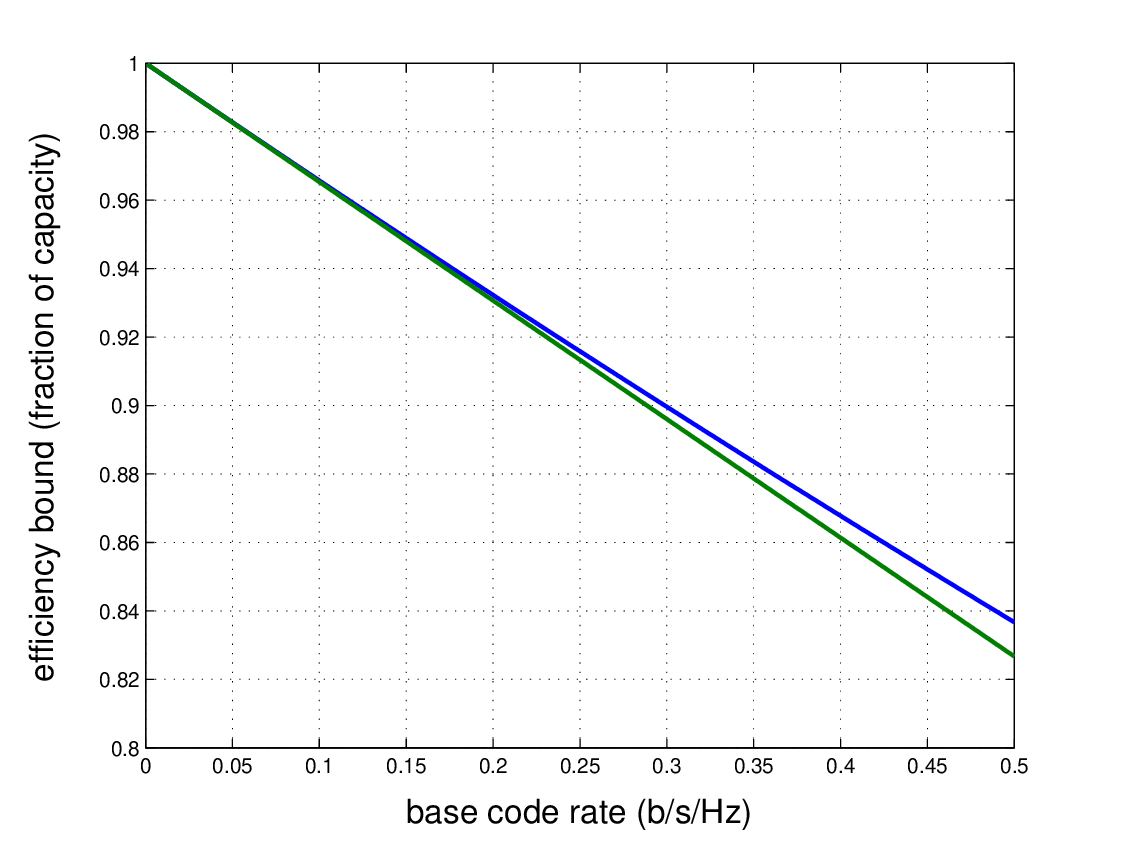}}
  \caption{Lower bound on efficiency of the near-perfect rateless code.
  The top and bottom curves are the middle and right-hand bounds of
  \protect\eqref{e:eff-bnds}, respectively. \label{f:eff}}
\end{figure}

In Fig.~\ref{f:eff}, the efficiency bounds \eqref{e:eff-bnds} are
plotted as a function of the base code rate $R''/L$.  As a practical
matter, our bound implies, for instance, that to obtain 90\%
efficiency requires a base code of rate of roughly $1/3$ bits per
complex symbol.  Note, too, that when the number of layers is
sufficiently large that the SNR per layer is low, a binary code may be
used instead of a Gaussian codebook, which may be convenient for
implementation.  For example, a code with rate $1/3$ bits per complex
symbol may be implemented using a rate-$1/6$ LDPC code with binary
antipodal signaling.

\section{Design and Implementation Issues}
\label{s:desimp}

In this section, we comment on some issues that arise in the
development and implementation of our rateless code constructions;
additional implementation issues are addressed in~\cite{sbw07}.


First, one consequence of our development of perfect rateless codes
for $M=L$ is that all layers must have the same rate $R/L$.  This does
not seem to be a serious limitation, as it allows a single base
codebook to serve as the template for all layers, which in turn
generally decreases the implementation complexity of the encoder and
decoder.  The codebooks $\mathcal{C}_1, \dots, \mathcal{C}_L$ used for
the $L$ layers should not be identical, however, for otherwise a naive
successive decoder might inadvertently swap messages from two layers
or face other difficulties that increase the probability of decoding
error.  A simple cure to this problem is to apply pseudorandom phase
scrambling to a single base codebook $\mathcal{C}$ to generate the
different codebooks needed for each layer.  Pseudorandom interleaving
would have a similar effect.

Second, it should be emphasized that a layered code designed with the
successive decoding constraint~\eqref{e:layerinf} can be decoded in a
variety of ways.  Because the undecoded layers act as colored noise,
an optimal decoder should take this into account, for example by using
a MMSE combiner on the received blocks $\{\rvv{y}_m\}$ as mentioned in
\secref{s:ldr}.  The MMSE combining weights change as each layer
is stripped off.  Alternatively, some or all of the layers could be
decoded jointly; this might make sense when the decoder for the base
codebook decoder is already iterative, and could potentially
accelerate convergence compared to a decoder that treats the layers
sequentially.

Third, a comparatively simple receiver is possible when all $M$
blocks have been received from a perfect rateless code in which $M=L$.
In this special case the linear combinations applied to the layers are
orthogonal, hence the optimal receiver can decode each layer
independently, without successive decoding.  This property is
advantageous in a multicasting scenario because it allows the
introduction of users with simplified receivers that function only at
certain rates, in this case the lowest supported one.


Finally, we note that with an ideal rateless code, \emph{every} prefix
of the code is a capacity-achieving code.  This corresponds to a
maximally dense set of SNR thresholds at which decoding can occur.  By
contrast, our focus in the paper has been on rateless codes that are
capacity-achieving only for prefixes whose lengths are an integer
multiple of the base block length.  The associated sparseness of SNR
thresholds can be undesirable in some applications, since when the
realized SNR is between thresholds, there is no guarantee that
capacity is achieved: the only realized rate promised by the
construction is that corresponding to the next lower SNR threshold.

However, as will be apparent from the simulations described in
\secref{s:sim}, performance is generally much better than this
pessimistic assessment.  In particular, partial blocks provide
essentially all the necessary redundancy to allow an appropriately
generalized decoder to operate as close to capacity as happens with
full blocks.

Nevertheless, when precise control over the performance at a dense set
of SNR thresholds is required, other approaches can be used.  For
example, when the target ceiling rate is $R$, we can use our rateless
code construction to design a code of ceiling rate $\kappa R$, where
$1\le\kappa\le M$, and have the decoder collect at least $\kappa$
blocks before attempting to decode.  With this approach, the
associated rate thresholds are $R, R\kappa/(\kappa+1),
R\kappa/(\kappa+2), \ldots, R\kappa/M$.  Hence, by choosing larger
values of $\kappa$, one can increase the density of SNR thresholds.

\section{Simulations}
\label{s:sim}

Implicit in our analysis is the use of perfect base codes and ideal
(maximum likelihood) decoding.  In this section, we present
simulations that further validate our
rateless code design with practical coding and decoding.

In our simulations, we use as our base code the turbo code specified
in the 3GPP LTE wireless standard \cite{LTE1,LTE2}.  This
parallel-concatenated convolutional code constructed from a pair of
8-state constituent encoders has a rate of $2/3$ bits per complex
symbol.  This code is used on conjunction with the iterative
turbo-decoding algorithm for which it was designed.

The base code is used in both 3- and 4-layer rateless constructions,
corresponding to ceiling rates of $R=2$ and $R=8/3$ b/s/Hz, respectively.
Moreover, there are a total of 6144 information bits per layer,
corresponding to a block length of $N=9216$ complex symbols.

Encoding proceeds as follows.  Since the base code is not ideal, it
has a bit-error rate that rolls off with the operating SNR.  Let
$\SNR_\circ(\epsilon)$ denote the SNR at which the base code achieves
a bit-error rate of $\epsilon$.  Then, using a definition analogous to
that used in \secref{s:eff}, the efficiency of the base code
is\footnote{One can equivalently measure the efficiency of the base
  code in terms of its gap to capacity at a particular target
  bit-error rate.   However, our chosen measure is more natural when
  relating the efficiency of the base code to the rateless code
  constructed from it.}
\begin{equation*} 
\eta_\circ(\epsilon) = \frac{R/L}{\log(1+\SNR_\circ(\epsilon))}.
\end{equation*}
Thus, in computing the gain matrix $\svv{G}$, we prescale the target
rate, replacing $R$ with $R/\eta_\circ(\epsilon)$.  Note that as a
result, $\svv{G}$ depends on the target rate and the base code
properties only.

For the particular base code used in the simulations, the efficiencies
are as given in Table~\ref{t:baseeff}.

\begin{table}
  \caption{Rate $2/3$ b/s/Hz 3GPP LTE Base Code Efficiencies
  \label{t:baseeff}}  
\begin{center}
\begin{tabular}{c|cccc}
&\multicolumn{4}{c}{Bit-Error Rate $\epsilon$}\\
& $10^{-2}$ & $10^{-3}$ & $10^{-4}$ & $10^{-5}$ \\ 
\hline
{\large\vphantom{M}}Efficiency $\eta_\circ$ & 88.9\% &  87.1\% &  85.7\% &  84.7\%
\end{tabular}
\end{center}
\end{table}

In our simulation, we decode not only from integer numbers of
redundancy blocks, but also from noninteger numbers, corresponding to
partial blocks.  In general, MMSE combining is applied on a
symbol-by-symbol basis, in conjunction with our usual successive
cancellation.  In particular, when the number of incremental
redundancy blocks $m$ is noninteger, then the MMSE combiner for the
first $N(m-\lfloor m \rfloor)$ symbols of the codeword in a given
layer $l$ is constructed from the submatrix $\svv{G}_{\lfloor m
  \rfloor+1,l}$, while the MMSE combiner for the remaining
$N(1+\lfloor m \rfloor - m)$ symbols of the codeword is constructed
from the submatrices $\svv{G}_{\lfloor m \rfloor,l}$.

Following combining (and cancellation), turbo decoding is applied to
the layer of interest, where the initial log-likelihood ratios are
calculated treating the symbols as corrupted by Gaussian noise with
variance determined by the effective SNR.  This effective SNR is
determined from the (reciprocal of the unbiased) mean-square error
resulting from MMSE combining, taking into account the successive
cancellation.  Thus, when $m$ is noninteger, the initial
log-likelihood ratios take on one value for the symbols in the
first part of the codeword, and a different value in the second part.

The overall efficiency $\eta$ of the resulting rateless code, i.e.,
the fraction of capacity at which it operates, is a function of the
number of incremental redundancy blocks $m$ (or equivalently the
realized SNR in the channel).  We calculate $\eta$ for the general
case where $m$ may be noninteger as follows.  First, for a given value
of $m$, the roll-off of the bit-error rate of the overall rateless
code as a function of the SNR can be generated, where for each SNR
value, the corresponding MMSE combiner with successive cancellation is
used.  As above, when $m$ is noninteger two MMSE combiners are
involved.  The resulting bit error rate is averaged over both the $N$
symbols within the codeword at every layer and the $L$ layers, so that
error propagation effects are taken into account.  We then let
$\SNR(m,\epsilon)$ denote the SNR at which the target bit-error rate
$\epsilon$ is attained for this particular value of $m$, from which
the efficiency of the rateless code is
\begin{equation} 
\eta(m,\epsilon) = \frac{R/m}{\log(1+\SNR(m,\epsilon))},
\end{equation}
where we have used a notion of efficiency consistent with earlier
definitions.

\begin{figure}[tbp]
\subfigure[3-layers, 3-blocks (rate range: $2/3$ to $2$
  b/s/Hz)]{\includegraphics[width=3.5in]{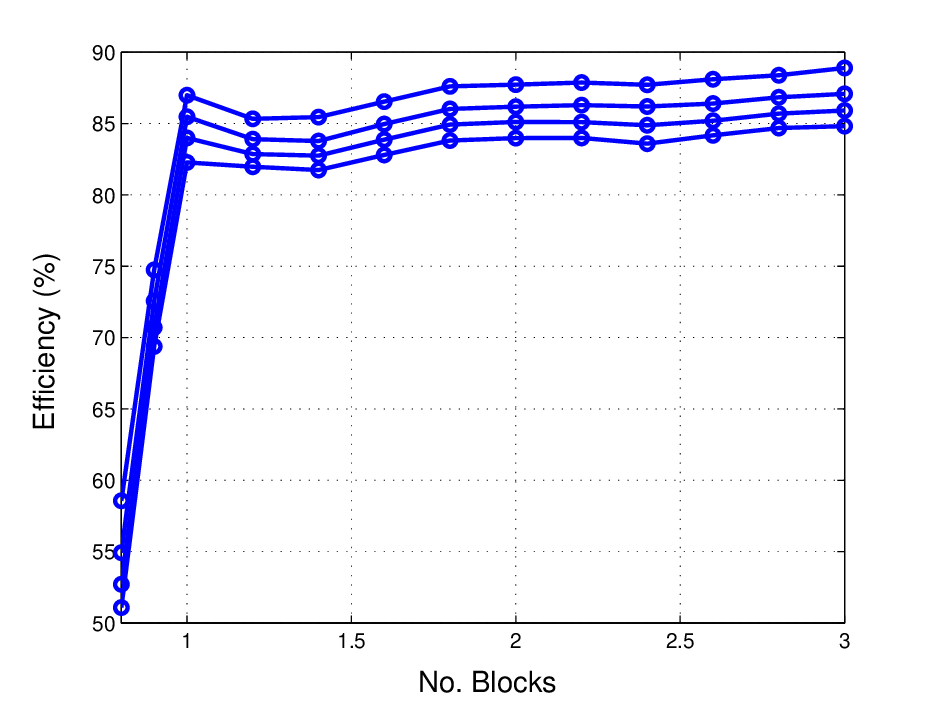}}\newline
\subfigure[4-layers, 4-blocks (rate range: $8/9$ to $8/3$
  b/s/Hz)]{\includegraphics[width=3.5in]{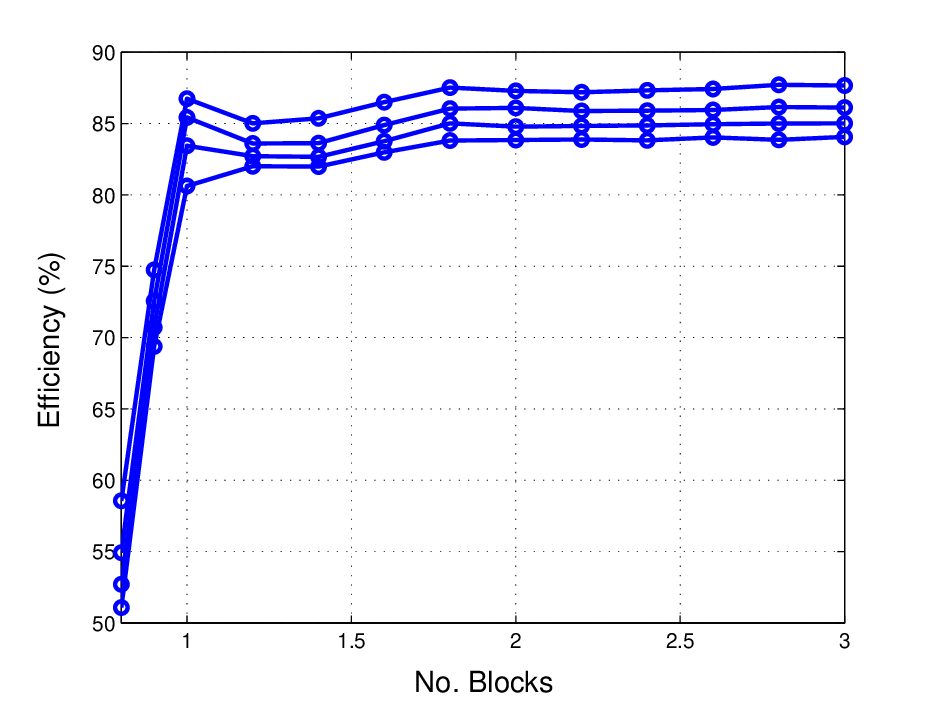}}
\caption{Practical efficiencies achieved using a rateless construction
  in conjunction with rate $2/3$ base code.  The successively lower
  curves correspond to target bit-error rates of $10^{-2}$, $10^{-3}$,
  $10^{-4}$, and $10^{-5}$, respectively.
\label{f:sim}}
\end{figure}

The resulting efficiency plots are depicted in Fig.~\ref{f:sim}.
Several features are noteworthy.  First, the efficiencies for
$m=1,2,\dots$ redundancy blocks are quite close to those of the base
code shown in Table~\ref{t:baseeff}; typically they are at most 2-3\%
lower.  This suggests, at least for codes with few layers, that losses
due to the rateless architecture itself, as well as the use of
iterative decoding in the face of non-Gaussian noise from undecoded layers,
are negligible in practice, and that good base
codes will yield good rateless codes.

Second, the efficiencies do not vary significantly with the number of
redundancy blocks $m$.  Moreover, even when partial redundancy blocks
are used, the efficiency does not deteriorate.  This suggests that our
rateless code constructions can operate over a much finer-grained set
of rates than our design prescribed.  However, it should be emphasized
that this holds only when at least one full redundancy block is used.
When less redundancy is used, Fig.~\ref{f:sim} shows that efficiency
falls off rapidly.

\section{Extensions to Time-Varying Channels}
\label{s:tvc}

The framework of this paper can be extended to time-varying channels
in a variety of ways.  As one example, the time-varying channel can be
viewed as an instance of parallel channels, and thus a solution can be
developed from a solution to the problem of rateless coding for
parallel channels.  Initial work in this direction is described in,
e.g., \cite{Shanechi,Shanechi-IZS,Shanechi-ISIT,Shanechi-GCOM}, though
much remains to be understood about the performance limits of various
constructions.  Another approach is based on the observation that 
feedback about the past channel state 
can significantly simplify the problem of encoding for future
transmissions \cite{Erez-IZS}.  It is this approach we describe here
as an illustration of potential.  In particular, we show that the
natural generalization of our architecture is perfect (i.e.,
capacity-achieving), enabling the message to be recovered with the
minimum possible number of blocks for the realized channel.

For the time-varying channel we consider, the
observations take the form
\begin{equation}
  \rvv{y}_m = \beta_m\rvv{x}_m + \rvv{z}_m,\quad m=1,2,\dots,
\label{e:tv-model}
\end{equation}
where the $\{\beta_m\}$ denote a sequence of complex channel gains.
The $\beta_m$ continue to be known a priori at the receiver but not at
the transmitter.

The encoder transmits a message $w$ by generating a sequence of
incremental redundancy blocks $\rvv{x}_1(w)$, $\rvv{x}_2(w)$, \dots.
The receiver accumulates sufficiently many received blocks
$\rvv{y}_1$, $\rvv{y}_2$, \dots to recover $w$.  Immediately following
the transmission of block $\rvv{x}_m$, the encoder is notified of
$\beta_m$.  Thus, knowledge of
$\beta_1,\dots,\beta_m$ can be used in the construction of the
redundancy block $\rvv{x}_{m+1}(w)$.

In this context, a \emph{perfect} rateless code is then one in which
capacity is achieved for any number $m=1,\dots,M$ of redundancy
blocks, i.e., whenever the (realized) channel gains are such that
\begin{equation}
R \le \sum_{m'=1}^m \log\left(1+\frac{P}{\sigma^2}|\beta_{m'}|^2\right),
\label{eq:perfect}
\end{equation}
the message can be recovered with high probability.

In this development, for values of $m$ such that the right side of
\eqref{eq:perfect} is less than $R$, it is convenient to define
\emph{target} channel gains $\alpha_{m+1}$ required for successful
decoding once block $m+1$ is obtained.  In particular, $\alpha_{m+1}$
is defined via
\begin{equation}
R = \log\left(1+\frac{P}{\sigma^2}\alpha_{m+1}^2\right) + \sum_{m'=1}^m
\log\left(1+\frac{P}{\sigma^2}|\beta_{m'}|^2\right),
\label{eq:al-def}
\end{equation}
whenever $\alpha_m >|\beta_m|$.

Generalizing our construction for the time-invariant case, we first
choose the range $M$, the ceiling rate $R$, the number of layers $L$,
and finally the associated base codebooks
$\mathcal{C}_1,\dots,\mathcal{C}_L$.  We assume a priori that the base
codebooks all have equal rate $R/L$.

As with our time-invariant construction, the redundancy blocks
$\rvv{x}_1,\dots,\rvv{x}_M$ take the form \eqref{e:rbasic}.  We
emphasize that the $m$\/th row of $\svv{G}$, which constitute the
weights used in constructing the $m$\/th redundancy block from the $L$
codewords, will in general be a function of the (realized) channel
gains $\beta_1,\dots,\beta_{m-1}$. Specifically, the $m$\/th row is
designed for the channel gain sequence $\{\beta_1,\dots,\beta_{m-1},\alpha_m\}$,
i.e., we substitute the target gain $\alpha_m$ for the (as yet unknown)
channel gain $\beta_m$. Finally, in addition to the layered code structure, we continue to
impose the constraint that the layered code be successively decodable.

Our aim is to select $\svv{G}$ so that the code is perfect as defined
earlier.  From the layered repetition encoding structure, we require
as in the time-invariant development that the rows of $\svv{G}$ be
orthogonal, while from the successive decoding constraint we have the
requirement [cf.\ \eqref{e:layerinf}] that
\begin{equation}
\frac{lR}{L} \le \log\det(\svv{I} + \frac{1}{\sigma^2}\svv{B}_m
  \svv{G}_{m,l}\svv{G}_{m,l}^\cgt \svv{B}_m^\cgt)
\label{e:tv-layerinf}
\end{equation}
for all $l=1,\dots,L$ and $m=1,\dots,M$, with
\begin{equation}
  \svv{B}_m =  \diag(\beta_1,\ldots,\beta_{m-1},\alpha_m).
\label{eq:B}
\end{equation}

With this model, in Appendix~\ref{app:tv-perfect} we construct in
closed form perfect rateless codes for the case of $M=2$ redundancy
blocks and $L=3$ layers for rates in the range $R<\log(2+\sqrt{5})
\approx 2.08$ bits per complex symbol.  This construction can be
viewed as the time-varying natural generalization of that in
\secref{s:example}.  Establishing the existence of perfect rateless
codes for larger values of $M$ and/or $L$ requires more effort.
However, following an approach analogous to that used in corresponding
development for the time-invariant case in \secref{s:existence}, we
shown in Appendix~\ref{app:tv-asymp} that in the limit of a large
number of layers $L$, asymptotically perfect codes for all values of
$M$ are possible.

\section{Concluding Remarks}
\label{s:conc}

In this paper, motivated by hybrid ARQ requirements in wireless and
related applications, our focus has been on the development of a
lossless framework for transforming a code good for the AWGN channel
at a single SNR into one good simultaneously at multiple SNRs.  There
are a variety of worthwhile directions for further research.

First, while beyond the scope of the present paper, a comparative
evaluation of methods described herein relative to, for example, those
described in \secref{s:back} is likely to reveal additional insight,
and uncover opportunities for further progress.

Second, while we have developed some preliminary results on the
extension of our framework to time-varying channels, clearly this is
just a beginning.  For example, when $M>2$, there is flexibility in the
problem formulation, and thus in how the available degrees of freedom
are allocated.  As another example, one could consider other
time-variation models, such as one that would allow $\beta$ to vary
deterministically so long as the pattern of variation is known in
advance.  Then, for one block the code would be designed for a gain of
$[\alpha_{1,1}]$, for two blocks the target gains would be
$[\alpha_{2,1}\ \alpha_{2,2}]$, for three blocks the gains would be
$[\alpha_{3,1}\ \alpha_{3,2}\ \alpha_{3,3}]$, and so on.  Still
another example would involve the development of solutions for
time-varying channels without requiring SNR feedback, either with or
without a stochastic model for $\beta$.

Other worthwhile directions include more fully developing rateless
constructions for the AWGN channel that allow decoding to begin at any
received block, and/or to exploit an arbitrary subset of the
subsequent blocks.  Initial efforts in this direction include the
faster-than-Nyquist constructions in \cite{Erez,Shanechi}, and the
diagonal subblock layering approach described in \cite{Shanechi}.

Beyond the single-input, single-output (SISO) channel, many
multiterminal and multiuser extensions are also of considerable
interest.  Examples of preliminary developments along these lines
include the rateless space-time code constructions in \cite{Erez-ST},
the rateless codes for multiple-access channels developed in
\cite{Niesen}, and the approaches to rateless coding for parallel
channels examined in
\cite{Shanechi,Shanechi-IZS,Shanechi-ISIT}.  Indeed, such
research may lead to efficient rateless orthogonal frequency-division
multiplexing (OFDM) systems and efficient rateless multi-input,
multi-output (MIMO) codes with wide-ranging applications.

Finally, extending the layered approach to rateless coding developed
in this paper beyond the Gaussian channel is also a potentially rich
direction for further research.  A notable example would be the binary
symmetric channel, where good rateless solutions remain elusive.
Preliminary work in this direction is described in
\cite{NarayananISIT}.

\appendices

\section{Perfect rateless solution for $L=M=3$} \label{a:rateless33}

Determining the set of solutions
\begin{equation}
\svv{G} = \begin{bmatrix}
  g_{11} & g_{12} & g_{13} \\
  g_{21} & g_{22} & g_{23} \\
  g_{31} & g_{32} & g_{33}
\end{bmatrix}
\end{equation}
to \eqref{e:layerinf} when $L=M=3$ as a function of the ceiling rate
$R$ is a matter of lengthy if routine algebra.

We begin by observing that any row or any column of $\svv{G}$ may be
multiplied by a common phasor without changing $\svv{G}\svv{G}^\cgt$.
Without loss of generality we may therefore take the first row and
first column of $\svv{G}$ to be real and positive.
Each $\svv{G}$ thus represents
a set of solutions $\svv{D}_1\svv{G}\svv{D}_2$, where $\svv{D}_1$ and
$\svv{D}_2$ are diagonal matrices in which the diagonal entries have
modulus 1.  The solutions in the set are equivalent for most
engineering purposes and we shall therefore not distinguish them
further.

We know that $\svv{G}$ must be a scaled unitary matrix, scaled so that
the row and column norms are $\sqrt{P}$.  Thus, if we somehow
determine the first two rows of $\svv{G}$, there is always a choice
for the third row: it's the unique vector orthogonal to the first two
rows which meets the power constraint and which has first component
real and positive.  Conversely, it's easy to see that any
appropriately scaled unitary matrix $\svv{G}$ that satisfies
\eqref{e:layerinf} for $m=1$ and $m=2$ (and all $l=1,2,3$)
necessarily satisfies \eqref{e:layerinf} for $m=3$.  We may therefore
without loss of generality restrict our attention to determining the
set of solutions to the first two rows of $\svv{G}$; the third row
comes ``for free'' from the constraint that $\svv{G}$ be a scaled
unitary matrix.

Assume, again without loss of generality, that $\alpha_1^2=1$ and
$\sigma^2=1$.  Via~\eqref{e:layerinf}, the first row of $\svv{G}$
(which controls the first redundancy block) must satisfy
\begin{align}
  R/3 &= \log (1 + g_{11}^2) \\
  2R/3 &= \log (1 + g_{11}^2 + g_{12}^2) \\
  3R/3 &= \log (1 + g_{11}^2 + g_{12}^2 + g_{13}^2)
\end{align}
and must also satisfy the power constraint
\begin{equation}
  P = g_{11}^2+g_{12}^2+g_{13}^2.
\end{equation}
Thus
\begin{equation*}
  P = 2^R - 1 = x^6-1
\end{equation*}
and
\begin{align}
  &{g_{11}^2  = 2^{R/3} - 1 = x^2-1,}\label{e:g11}\\
  &{g_{12}^2 = 2^{R/3}(2^{R/3}-1) = x^2(x^2-1),}\label{e:g12} \\
  &{g_{13}^2 = 2^{2R/3}(2^{R/3}-1) = x^4(x^2-1),} \label{e:g13}
\end{align}
where for convenience we have introduced the change of variables
$x=2^{R/6}$.

The first column of $\svv{G}$ (which controls the first layer of each
redundancy block) is also straightforward.  Via~\eqref{e:alpha} with
$m=2$ and $m=3$, we have
\begin{align}
  \alpha_2^2 &= \frac{1}{x^3+1}, \label{e:alpha2}\\
  \alpha_3^2 &= \frac{1}{x^4+x^2+ 1}.
\end{align}
Using~\eqref{e:layerinf} for $l=1$ and $m=2$ yields
\begin{equation}
  R/3 = \log(1 + \alpha_2^2(g_{11}^2+g_{21}^2)).
\label{eq:r3}
\end{equation}
Substituting the previously computed expressions \eqref{e:g11} and
\eqref{e:alpha2} for $g_{11}^2$ and
$\alpha_2^2$ into \eqref{eq:r3} and solving for $g_{21}$ yields
\begin{equation}
  {g_{21}^2 = x^3(x^2-1).}
\label{e:g21}
\end{equation}

To solve for the second row of $\svv{G}$ we use \eqref{e:layerinf}
with $m=l=2$ together with the requirement that the first and second
rows be orthogonal.  It is useful at this stage to switch to
polar coordinates, i.e., $g_{22} = |g_{22}|e^{j\theta_1}$ and $g_{23} =
|g_{23}|e^{j\theta_2}$.

Orthogonality of the first and second rows means that
\begin{equation}
  0 = g_{11}g_{21} + g_{12}|g_{22}|e^{j\theta_1} +
  g_{13}|g_{23}|e^{j\theta_2}. \label{e:ortho12} 
\end{equation}
Complex conjugation is not needed here because the first row is real.
The three terms in the above expression may be viewed as the legs of a triangle,
so by the law of cosines
\begin{equation}
  2g_{11}g_{21}g_{12}|g_{22}|\cos\theta_1 = g_{13}^2|g_{23}|^2-g_{11}^2g_{21}^2-g_{12}^2|g_{22}|^2. 
\label{e:ti-uglytheta}
\end{equation}

We now use~\eqref{e:layerinf} with $m=l=2$ to infer that
\begin{equation}
  2^{2R/3} = x^4=\det(\svv{I} +
  \alpha_2^2\svv{G}_{2,2}\svv{G}_{2,2}^\cgt).
  \label{e:logdet22a}
\end{equation}
To expand this expression, we compute
\begin{equation}
  \svv{G}_{2,2}\svv{G}_{2,2}^\cgt
  =
  \begin{bmatrix}
    g_{11}^2+g_{12}^2 & g_{11}g_{21}+g_{12}|g_{22}|e^{-j\theta_1}\\
    (\ast) & g_{21}^2+|g_{22}|^2
  \end{bmatrix},
\end{equation}
where $(\ast)$ is the complex conjugate of the upper right entry, from
which we find
\begin{multline}
  \det(\svv{I} + \alpha_2^2\svv{G}_{2,2}\svv{G}_{2,2}^\cgt)
   = \\
   \alpha_2^4(g_{11}^2|g_{22}|^2+g_{12}^2g_{21}^2-2g_{11}g_{21}g_{12}|g_{22}|\cos\theta_1)\\
   {}+\alpha_2^2(g_{11}^2+g_{12}^2+g_{21}^2+|g_{22}|^2)+1.
     \label{e:ti-logdet22b}
\end{multline}
Substituting \eqref{e:ti-uglytheta} into \eqref{e:ti-logdet22b} to eliminate the cosine term
and using \eqref{e:logdet22a} yields
\begin{align}
  x^4 &= \alpha_2^4 (g_{11}^2|g_{22}|^2+g_{12}^2g_{21}^2-g_{13}^2|g_{23}|^2 \notag\\
  &\qquad\qquad\qquad {}+g_{11}^2g_{21}^2+g_{12}^2|g_{22}|^2) \notag\\
&\quad {}+\alpha_2^2(g_{11}^2+g_{12}^2+g_{21}^2+|g_{22}|^2)+1.
\end{align}

Finally, substituting the expressions for $g_{11}^2$, $g_{12}^2$,
$g_{13}^2$, $g_{21}^2$, and $\alpha_2^2$ computed above, using the power
constraint
\begin{equation}
  |g_{23}|^2 = P - |g_{22}|^2 - g_{21}^2, \label{e:powerrow2}
\end{equation}
solving for $|g_{22}|^2$, and simplifying, we arrive at
\begin{equation}
  {|g_{22}|^2 = (x^5+1)(x-1).}
\end{equation}
The power constraint \eqref{e:powerrow2} then immediately yields
\begin{equation}
  {|g_{23}|^2 = x(x^2-1).}
\end{equation}

The squared modulus of the entries of the last row of $\svv{G}$
follow immediately from the norm constraint on the columns:
\begin{equation}
  {g_{31}^2 = P - g_{21}^2+g_{11}^2 =  x^2(x^2-x+1)(x^2-1).}
\end{equation}
\begin{equation}
  {|g_{32}|^2 = P - g_{22}^2 - g_{12}^2 = x(x^3+1)(x-1)}
\end{equation}
and
\begin{equation}
  {|g_{33}|^2 = P - g_{23}^2 - g_{13}^2 = (x^3+1)(x-1).}
\end{equation}
This completes the calculation of the squared modulus of the entries
of $\svv{G}$.  In summary, we have shown that $\svv{G}$ has the form
\begin{multline}
  \svv{G} = \sqrt{x-1} \cdot{} \\
  \begin{bmatrix}
    \sqrt{x+1} & \sqrt{x^2(x+1)} & \sqrt{x^4(x+1)} \\
    \sqrt{x^3(x+1)} & e^{j\theta_1}\sqrt{x^5+1} & e^{j\theta_2}\sqrt{x(x+1)} \\
    \sqrt{x^2(x^3+1)} & e^{j\theta_3}\sqrt{x(x^3+1)} & e^{j\theta_4}\sqrt{x^3+1}
  \end{bmatrix} \label{e:rateless33}
\end{multline}
where $x=2^{R/6}$.

We must now establish the existence of suitable
$\theta_1,\dots,\theta_4$.  To resolve this question it suffices to
consider the consequences of the orthogonality constraint
\eqref{e:ortho12} on $\theta_1$ and $\theta_2$.  As remarked at the
start of this section, the last row of $\svv{G}$ and hence $\theta_3$
and $\theta_4$ come ``for free'' once we have the first two rows of
$\svv{G}$.

Substituting the expressions for $|g_{ml}|^2$ determined above into
\eqref{e:ortho12} and canceling common terms yields
\begin{equation}
  0 = \sqrt{x} + e^{j\theta_1}\sqrt{x^4-x^3+x^2-x+1} +
  e^{j\theta_2}\sqrt{x^3}.
  \label{e:triangle}
\end{equation}
The right-hand side is a sum of three phasors of predetermined
magnitude, two of which can be freely adjusted in phase.  In geometric
terms, the equation has a solution if we can arrange the three complex
phasors into a triangle, which is possible if and only if the longest
side of the triangle is no longer than the sum of the lengths of the
shorter sides.  The resulting triangle is unique (up to complex
conjugation of all the phasors).  Now, the middle term of
\eqref{e:triangle} grows faster in $x$ than the others, so for large
$x$ we cannot possibly construct the desired triangle.  A necessary
condition for a solution is thus
\begin{equation}
  \sqrt{x} + \sqrt{x^3} \ge \sqrt{x^4-x^3+x^2-x+1},
\end{equation}
where equality can be shown (after some manipulation) to hold at the
largest root of $x^2-x+1$, i.e., at $x=(3+\sqrt{5})/2$, or
equivalently $R=6\log_2 x = 6\log_2(3+\sqrt{5}) - 6$.  It becomes
evident by numerically plotting the quantities involved that this
necessary condition is also sufficient, i.e., a unique solution to
\eqref{e:triangle} exists for all values of $x$ in the range $1<x\le
(3+\sqrt{5})/2$ and no others.  Establishing this fact algebraically
is an unrewarding though straightforward exercise.

A relatively compact formula for $\theta_1$ may be found by applying
the law of cosines to \eqref{e:triangle}:
\begin{equation}
  \cos(\pi-\theta_1) = \frac{x^4-2x^3+x^2+
    1}{2\sqrt{x(x^4-x^3+x^2-x+1)}}.
\end{equation}
Similar formulas may be derived for $\theta_2$, $\theta_3$, and $\theta_4$.

\section{Power Allocation} \label{a:power}

The power allocation satisfying the property \eqref{e:pa-cond} can be
obtained as the solution to a different but closely related rateless
code optimization problem.  Specifically, let us retain the block
structuring and layering of the code of \secref{s:enc}, but
instead of using repetition and dithering in the construction, let us
consider a code where the codebooks in a given layer are
\emph{independent} from block to block.  While such a code is still
successively decodable, it does not retain other characteristics that
make decoding possible with low complexity.  However, the complexity
characteristic is not of interest.  What does matter to us is that
the per-layer, per-block SNRs that result from a particular power
allocation will be identical to those of the code of \secref{s:enc}
for the same power allocation.  Thus, in tailoring our code in this
Appendix to meet \eqref{e:pa-cond}, we simultaneously ensure our code
of \secref{s:enc} will as well.

We begin by recalling a useful property of layered codes in general
that we will apply.  Consider an AWGN channel with gain $\beta$ and
noise $\rvs{z}$ of variance $\sigma^2$, and consider an $L$-layer block
code that is successively decodable. If the constituent codes are
capacity-achieving i.i.d.\ Gaussian codes, and MMSE successive
cancellation is used, then the overall code will be capacity
achieving.  More specifically, for any choice of powers $p_l$ for
layers $l=1,2,\dots,L$ that sum to the power constraint $P$, the
associated rates $I_l$ for these layers will sum to the corresponding
capacity $\log(1+|\beta|^2P/\sigma^2)$.  Equivalently, for any choice
of rates $I_l$ that sum to capacity, the associated powers $p_l$ will
sum to the corresponding power constraint.  In this latter case, any
rate allocation that yield powers that are all nonnegative is a valid
one.

To see this, let the relevant codebooks for the layers be
$\tilde{\mathcal{C}}_1,\dots,\tilde{\mathcal{C}}_L$, and let the
overall codeword be denoted
\begin{equation}
  \tilde{\rvs{c}} = \tilde{\rvs{c}}_1 + \dots + \tilde{\rvs{c}}_L,
\end{equation}
where the $\tilde{\rvs{c}}_l\in\tilde{\mathcal{C}}_l$ are
independently selected codewords drawn for each layer.  The overall
code rate is the sum of the rates of the individual codes.  The
overall power of the code is $P = p_1+\dots+p_L$.

From the mutual information decomposition
\begin{equation} 
I(\tilde{\rvs{c}};\rvs{y}) = \sum_{l=1}^L I_l
\label{e:chain}
\end{equation}
where
\begin{equation*} 
I_l =
I(\tilde{\rvs{c}}_l;\tilde{\rvs{c}}_1+\dots+\tilde{\rvs{c}}_L+\rvs{z}
\mid \tilde{\rvs{c}}_{l+1}^{L}),  
\end{equation*}
with
$\tilde{\rvs{c}}_{l+1}^{L}=(\tilde{\rvs{c}}_{l+1},\tilde{\rvs{c}}_{l+2},\dots,\tilde{\rvs{c}}_{L})$,
we see that the overall codebook power constraint $P$ can be met by
apportioning power to layers in any way desired, so long as
$p_1+\dots+p_L=P$.  Since the undecoded layers are treated as noise,
the maximum codebook rate for the $l$\/th layer is then
\begin{equation}
  I_l = \log(1+\SNR_l) \label{e:Il}
\end{equation}
where
\begin{equation}
  \SNR_l = \frac{|\beta|^2 p_l}{|\beta|^2 p_1+|\beta|^2
  p_2+\dots+|\beta|^2
  p_{l-1} + \sigma^2} \label{e:SNR}
\end{equation}
is the effective SNR when decoding the $l$\/th layer.  Straightforward
algebra, which amounts to a special-case recalculation of
\eqref{e:chain}, confirms that $I_1+\dots+I_L=\log(1+|\beta|^2
P/\sigma^2)$ for any selection of powers $\{p_l\}$.

Alternatively, instead of selecting per-layer powers and computing
corresponding rates, one can select per-layer rates and compute the
corresponding powers.  The rates $\{I_l\}$ for each level may be set
in any way desired so long as the total rate $I_1+\dots+I_L$ does not
exceed the channel capacity $\log(1+|\beta|^2 P/\sigma^2)$.  The
required powers $\{p_l\}$ may then be found using \eqref{e:Il} and
\eqref{e:SNR} recursively for $l=1,\dots,L$.  There is no need to
verify the power constraint: it follows from \eqref{e:chain} that the
powers computed in this way sum to $P$.  Thus it remains only to check
that the $\{p_l\}$ are all nonnegative to ensure that the rate
allocation is a valid one.

We now apply this insight to our rateless context.  The target ceiling
rate for our rateless code is $R$, and, as before, $\alpha_m$,
$m=1,2,\dots$, denotes the threshold channel gains as obtained via
\eqref{e:alpha}.  

Comparing \eqref{e:pa-cond} with \eqref{e:Il}
and \eqref{e:SNR} reveals that \eqref{e:pa-cond} can be rewritten as
\begin{equation} 
R_l = \sum_{m'=1}^m I_{m',l} (\alpha_m),
\label{e:pa-interp}
\end{equation}
for all $l=1,2,\dots,L$ and $m=1,2,\ldots$, where
\begin{equation} 
R_l = R/L
\label{e:Rl-choice}
\end{equation}
and $I_{m',l}(\alpha_m)$ is the mutual information in layer $l$ from
block $m'$ when the realized channel gain is $\alpha_m$.  Thus,
meeting \eqref{e:pa-cond} is equivalent to finding powers $p_{m',l}$
for each code block $m'$ and layer $l$ so that for the given rate
allocation $R_l$ (a) the powers are nonnegative, (b) the power
constraint is met, and (c) when the channel gain is $\alpha_m$, the
mutual information accumulated at the $l$\/th layer after receiving
code blocks $1,2,\dots,m$ equals $R_l$.

Since the power constraint is automatically satisfied by any
assignment of powers that achieves the target rates, it suffices to
establish that \eqref{e:pa-interp} has a solution with nonnegative
per-layer powers.

The solution exists and is unique, as can be established by induction
on $m$.  Specifically, for $m=1$ the rateless code is an ordinary
layered code and the powers $p_{1,1},\dots,p_{1,L}$ may be computed
recursively from [cf.\ \eqref{e:pa-interp}]
\begin{equation} 
  R_l = \sum_{m'=1}^m \log(1+\SNR_{m',l}(\alpha_m)),
\label{e:Rl}
\end{equation}
with $\SNR_{m,l}(\alpha_m)$ as given in \eqref{e:SNRml} for
$l=1,\dots,L$.

For the induction hypothesis, assume we have a power assignment for
the first $m$ blocks that satisfies \eqref{e:Rl}.  To find the
power assignment for the $(m+1)$\/st block, observe that when the
channel gain decreases from $\alpha_m$ to $\alpha_{m+1}$ the
per-layer mutual information of every block decreases.  A
nonnegative power must be assigned to every layer in the $(m+1)$\/st
code block to compensate for the shortfall.

The mutual information shortfall in the $l$\/th layer is
\begin{equation}
  \Delta_{m+1,l} = R_l - \sum_{m'=1}^m \log(1+\SNR_{m',l}(\alpha_{m+1})),
  \label{e:dI}
\end{equation}
and the power $p_{m+1,l}$ needed to make up for this shortfall is
the solution to
\begin{equation}
  \Delta_{m+1,l} = \log(1+\SNR_{m+1,l}(\alpha_{m+1})),
\end{equation}
viz.,
\begin{multline}
  p_{m+1,l} = (2^{2 \Delta_{m+1,l}}-1)\\
  {}\cdot (p_{m+1,1}+\dots+p_{m+1,l-1}+\frac{\sigma_{m+1}^2}{\alpha_{m+1}^2}).
  \label{e:Pml}
\end{multline}
This completes the induction.  Perhaps counter to intuition, even if
the per-layer rates $R_1,\dots,R_L$ are set equal, the per-layer
shortfalls $\Delta_{m+1,1},\dots,\Delta_{m+1,L}$ will not be equal.
Thus, within a layer the effective SNR and mutual information will
vary from block to block.

Eqs.~\eqref{e:dI} and \eqref{e:Pml} are easily evaluated numerically.
An example is given in Table~\ref{t:powers}.\footnote{If one were
aiming to use a rateless code of the type described in
\secref{s:existence} in practice, in calculating a power allocation
one should take into account the gap to capacity of the particular
base code being used.  Details of this procedure for the case of
perfect rateless codes are given as part of the description of the
simulations in Section~\ref{s:sim}.  For the case of near perfect
codes, the corresponding procedure is described in \cite{sbw07}.}

\begin{table}
\caption{Per-layer power assignments $p_{m,l}$ and 
channel gain thresholds $\alpha_m$
  for the initial blocks of an
  $L=4$ layer rateless code with total power $P=255$, noise
  variance $\sigma^2=1$, and per-layer
  rate $R/L=1$ b/s/Hz. \label{t:powers}}  
\begin{center}
    \begin{tabular}{r|rrrrr}
      & $m=1$ & $m=2$ & $m=3$ & $m=4$ & $m=5$\\\hline
      gain (dB) & 0.00 & -12.30 & -16.78 & -19.29 & -20.99\\\hline
      $l=1$ &    3.00  &40.80&  48.98 & 55.77 & 58.79\\
      $l=2$ &    12.00  &86.70&  61.21 & 60.58 & 61.65\\
      $l=3$ &    48.00  &86.70&  81.32 & 71.48 & 67.50\\
      $l=4$ &   192.00  &40.80&  63.48 & 67.16 & 67.06\\
    \end{tabular}
  \end{center}
\end{table}

Finally, since this result holds regardless of the choice of the
constituent $R_l$, it will hold for the particular choice
\eqref{e:Rl-choice}, whence \eqref{e:pa-cond}.

\section{Perfect $L=3$, $M=2$ Rateless Solution for Time-Varying Channel}
\label{app:tv-perfect}

As the simplest example, for the case of $M=2$ redundancy blocks and
$L=3$ layers the constraints \eqref{e:tv-layerinf} can be met, i.e., a perfect rateless
code is possible provided $R$ is not too large.

In this case, we determine our gain matrix
\begin{equation}
  \svv{G} = \begin{bmatrix}
    g_{11} & g_{12} & g_{13}\\
    g_{21} & g_{22} & g_{23}
  \end{bmatrix}
\end{equation}
as a function of the ceiling rate $R$, where the second row also
depends on the realized channel gain $\beta_1$ experienced
by the first incremental redundancy block.

As in the time-invariant case, we may without loss of generality take the first
row and column to be real and nonnegative.  Assume, also without loss
of generality, that $\alpha_1=1$ and $\sigma^2=1$.  Then the first row of
$\svv{G}$, which corresponds to the first redundancy block, is
computed exactly as in the time-invariant case.  In particular, from
\eqref{e:tv-layerinf} with $m=1$, it must satisfy
\begin{align}
  R/3 &= \log (1 + g_{11}^2) \\
  2R/3 &= \log (1 + g_{11}^2 + g_{12}^2) \\
  3R/3 &= \log (1 + g_{11}^2 + g_{12}^2 + g_{13}^2)
\end{align}
together with the power constraint
\begin{equation}
  P = g_{11}^2+g_{12}^2+g_{13}^2.
\end{equation}
Thus, with $x\triangleq 2^{R/6}$, we have
\begin{equation}
  P = 2^R - 1 = x^6-1
\end{equation}
and
\begin{align}
  &g_{11}^2  = 2^{R/3} - 1 = x^2-1,\label{e:tv-g11}\\
  &g_{12}^2 = 2^{R/3}(2^{R/3}-1) = x^2(x^2-1),\label{e:tv-g12} \\
  &g_{13}^2 = 2^{2R/3}(2^{R/3}-1) = x^4(x^2-1).
\end{align}

The derivation now departs from the time-invariant case.  Recall that
$\beta_1$ is the realized channel gain for the first block.  A second
redundancy block is thus needed when $|\beta_1|<\alpha_1$.  The
target gain $\alpha_2$ is the solution to [cf.\ \eqref{eq:al-def}]
\begin{equation}
  R = \log(1+P|\beta_1|^2) + \log(1+P\alpha_2^2),
\end{equation}
which is
\begin{equation}
  \alpha_2^2 = \frac{1-|\beta_1|^2}{1+P|\beta_1|^2}.\label{e:tv-alpha2}
\end{equation}

Using \eqref{e:tv-layerinf} for $m=2$ and $l=1$ yields
\begin{equation}
  R/3 = \log(1+|\beta_1|^2g_{11}^2 + \alpha_2^2g_{21}^2).
\end{equation}
Substituting the previously computed expressions \eqref{e:tv-g11} and
\eqref{e:tv-alpha2} for $g_{11}^2$ and $\alpha_2^2$ and solving for
$g_{21}$ yields
\begin{equation}
  g_{21}^2 = (x^2-1)(1+P|\beta_1|^2).
\end{equation}

As in the time-invariant case, to solve for the rest of the second row
of $\svv{G}$ we use \eqref{e:tv-layerinf} with $m=l=2$ together with the
requirement that the first and second rows be orthogonal.  It is
useful at this stage to switch to polar coordinates, i.e., $g_{22} =
|g_{22}|e^{j\theta_1}$ and $g_{23} = |g_{23}|e^{j\theta_2}$.

Orthogonality of the first and second rows means that
\begin{equation}
  0 = g_{11}g_{21} + g_{12}|g_{22}|e^{j\theta_1} +
  g_{13}|g_{23}|e^{j\theta_2}. \label{e:tv-ortho12} 
\end{equation}
The three terms in the above expression may be viewed as the legs of a
triangle, so by the law of cosines
\begin{equation}
  2g_{11}g_{21}g_{12}|g_{22}|\cos\theta_1 = g_{13}^2|g_{23}|^2-g_{11}^2g_{21}^2-g_{12}^2|g_{22}|^2. 
\label{e:tv-uglytheta}
\end{equation}
We now use~\eqref{e:tv-layerinf} with $m=l=2$ to infer that
\begin{equation}
  2^{2R/3} = x^4=\det(\svv{I} +
  \text{diag}\{|\beta_1|^2|,\alpha_2^2\}\svv{G}_{2,2}\svv{G}_{2,2}^\cgt).
  \label{e:tv-logdet22a}
\end{equation}
To expand this expression, we compute
\begin{equation}
  \svv{G}_{2,2}\svv{G}_{2,2}^\cgt
  =
  \begin{bmatrix}
    g_{11}^2+g_{12}^2 & g_{11}g_{21}+g_{12}|g_{22}|e^{-j\theta_1}\\
    (\ast) & g_{21}^2+|g_{22}|^2
  \end{bmatrix},
\end{equation}
where $(\ast)$ is the complex conjugate of the upper right entry, from
which we find
\begin{multline}
  \det(\svv{I} + \text{diag}\{|\beta_1|^2|,\alpha_2^2\}\svv{G}_{2,2}\svv{G}_{2,2}^\cgt)
   = \\
   |\beta_1|^2\alpha_2^2(g_{11}^2|g_{22}|^2+g_{12}^2g_{21}^2-2g_{11}g_{21}g_{12}|g_{22}|\cos\theta_1)\\
   {}+|\beta_1|^2(g_{11}^2+g_{12}^2)+\alpha_2^2(g_{21}^2+|g_{22}|^2)+1.
     \label{e:tv-logdet22b}
\end{multline}
Substituting \eqref{e:tv-uglytheta} into \eqref{e:tv-logdet22b} and using
\eqref{e:tv-logdet22a} yields
\begin{align}
  x^4 &= |\beta_1|^2\alpha_2^2 \notag\\
&\quad {}\cdot
  (g_{11}^2|g_{22}|^2+g_{12}^2g_{21}^2-g_{13}^2|g_{23}|^2+g_{11}^2g_{21}^2+g_{12}^2|g_{22}|^2)\\
&\qquad\quad {}+|\beta_1|^2(g_{11}^2+g_{12}^2)+\alpha_2^2(g_{21}^2+|g_{22}|^2)+1.
\end{align}

Finally, substituting the expressions for $g_{11}^2$, $g_{12}^2$,
$g_{13}^2$, $g_{21}^2$, and $\alpha_2^2$ computed above, using the power
constraint
\begin{equation}
  |g_{23}|^2 = P - |g_{22}|^2 - g_{21}^2, 
\end{equation}
solving for $|g_{22}|^2$, and simplifying terms, we arrive at
\begin{multline}
  |g_{22}|^2 = \frac{x^2-1}{1+(x^6-1)|\beta_1|^2} \\
  {}\cdot \Bigl(x^2+|\beta_1|^2(x^{10}+x^8-x^6-x^4-x^2+1)\\
  {}-|\beta_1|^4(x^6-1)^2\Bigr) .
\end{multline}

Evidently, a necessary condition for the existence of a solution for
$\svv{G}$ is that $g_{21}^2+|g_{22}|^2< P$.  It can be shown that the
sum of the powers on the first two layers is maximized when
$|\beta_1|\to 1$, and then the necessary condition simplifies to
\begin{equation}
  2^{R+1} - 2^{2R/3+1} < 2^R - 1,
\end{equation}
which may be shown to hold for all $R<\log(2+\sqrt{5}) \approx 2.08$
bits per complex symbol.

The final step---a straightforward exercise, the details of which we
omit---is to apply the triangle inequality to \eqref{e:tv-ortho12} to
prove that the required triangle exists, and thus the phases
$\theta_1$ and $\theta_2$.

\section{Near-Perfect Rateless Codes for Time-Varying Channels}
\label{app:tv-asymp}

Our construction is a slight generalization of the corresponding
construction in \secref{s:existence} for time-invariant
channels.  First, we fix $M$, $R$, $L$, and the associated codebooks
$\mathcal{C}_l,\dots,\mathcal{C}_L$ each of rate $R'/L$ for some
$R'<R$ to be determined. Using $\rvs{c}_l(n)$ and $\rvs{x}_m(n)$ to
denote the $n$\/th elements of codeword $\rvv{c}_l$ and redundancy
block $\rvv{x}_m$, respectively, we again have \eqref{e:rbasic2}.

\subsection*{Power Allocation}
\label{sec:tv-power}

As in the corresponding development for the time-invariant case in
\secref{a:power}, a suitable power allocation for our construction is
obtained as that which is optimum for a slightly different
construction, which we now develop.  In this section, different
(independent) codebooks are used for different redundancy blocks, and
we take $\svv{G}(n)$ to be independent of $n$, so that
$\svv{G}(n)=\svv{P}$, where $\svv{P}$ is as given in \eqref{e:P}.

The mutual information in the $l$\/th layer of the $m$\/th block is then
\begin{equation}
I_{m,l}=\log(1+\SNR_{m,l}(\beta_m)).
\label{eq:Cml1}
\end{equation}
where
\begin{equation}
  \SNR_{m,l}(\beta_m) = \frac{|\beta_m|^2
  p_{m,l}}{|\beta_m|^2(p_{m,1}+\dots+p_{m,l-1}) + 1}.
\label{e:tv-SNRml}
\end{equation}
is the associated per-layer SNR experienced during successive
decoding.

We now obtain the elements of $\svv{P}$ recursively. We proceed from
the first block $m=1$ to block $M$, where in each block $m$ we start
by determining $P_{m,1}$ and proceed up through $P_{m,L}$. By
definition of $\alpha_1$, we have
\begin{equation*} 
\log \left( 1+\alpha_1^2 \sum_{l=1}^{L}P_{1,l} \right)=R.
\end{equation*}
Viewing the layering as superposition coding for a multi-access
channel, it is clear that any rate vector is achievable as long as its
sum-rate is $R$. We may therefore obtain an equal rate per layer by
taking $P_{1,1},\ldots,P_{1,L}$ such that
\begin{equation}
\log(1+P_{1,l}\alpha_1^2) = R/L,\quad l=1,\dots,L.
\end{equation}
Upon receiving knowledge of $|\beta_1|$ we proceed to determine the
power allocation for block $m=2$. More generally, suppose the power
allocations through block $m-1$ have been determined and we have now
acquired channel state knowledge through $\beta_{m-1}$. To determine
the allocation for block $m$, we first compute the mutual information
shortfall in layer $l$ as
\begin{equation}
\Delta_{m,l} = \frac{R}{L} - \sum_{m'=1}^{m-1} \log(1+\SNR_{m',l}(\beta_{m'})).
 \label{eq:Dml}
\end{equation}
By the induction hypothesis, had the realized channel gain been $|\beta_{m-1}|=\alpha_{m-1}$,
then $\Delta_{m,l}$ would be zero for all $l=1,\ldots,L$. Now since we have $|\beta_{m-1}| <
\alpha_{m-1}$, clearly the shortfall is positive for all layers. Also, by definition of
$\alpha_m$, we also have
\begin{equation}
\Delta_m = \sum_{l=1}^{L} \Delta_{m,l} = \log(1+P\alpha_m^2).
\end{equation}

We then solve for $p_{m,1},\dots,p_{m,L}$, in order, via
\begin{equation}
\log(1+\SNR_{m,l}(\alpha_m)) = \Delta_{m,l}.
\end{equation}

The resulting power allocation ensures that the aggregate mutual
information per layer is at least $R/L$ if $|\beta_m|> \alpha_m$
when i.i.d. Gaussian codebooks for all layers and blocks.  However, we
wish to use the same set of $L$ codebooks for all redundancy blocks,
to keep decoding complexity low.  We return to this problem next, but
in doing so will exploit this power allocation.

\subsection*{Encoding}
\label{s:tv-enc}

In our construction we restrict our attention to an encoding of the
form described in \secref{s:enc}.  In particular, the $\rvv{G}(n)$ are
of the form \eqref{e:PD_form} with \eqref{e:P} and \eqref{e:D}, with
the $\rvs{d}_{m,l}(n)$ all i.i.d.\ random variables in $m$, $l$, and
$n$, and drawn independently of all other random variables, including
noises, messages, and codebooks.  As before, it is sufficient for
$\rvs{d}_{m,l}(n)$ to take on only values $\pm1$, and with equal
probability.

\subsection*{Decoding}
\label{s:tv-dec}

Decoding proceeds in a manner analogous to that described in
\secref{s:dec} for the time-invariant case.  In particular, since
$\rvv{G}(n)$ is drawn i.i.d., the overall channel is i.i.d., and thus
we may express the channel model in terms of an arbitrary individual
element in the block.  Specifically, assume that the channel gain for
block $m$ is the minimal required $\beta_m=\alpha_m$, then our
received symbol can be expressed as [cf.\ \eqref{e:rx}]
\begin{equation*}
\rvv{y} =
\begin{bmatrix}
\rvs{y}_1\\ \vdots\\ \rvs{y}_m
  \end{bmatrix}
   = \svv{B}_m \rvv{G}
  \begin{bmatrix}
   \rvs{c}_1\\ \vdots\\ \rvs{c}_L
  \end{bmatrix}
  +
  \begin{bmatrix}
    \rvs{z}_1\\ \vdots\\ \rvs{z}_m
  \end{bmatrix},
\end{equation*}
where $\rvv{G}= \svv{P}\odot \rvv{D}$, with $\rvv{G}$ denoting the
arbitrary element in the sequence $\rvv{G}(n)$, and where
$\rvs{y}_{m'}$ is the corresponding received symbol from redundancy block
$m'$ (and similarly for $\rvs{c}_{m'}$, $\rvs{z}_{m'}$, $\rvv{D}$).

As in the time-invariant case, it is sufficient to employ successive
cancellation decoding with simple maximal ratio combining (MRC) of the
redundancy blocks.  In this case, the effective SNR at which this
$l$\/th layer is decoded from $m$ blocks via such MRC decoding is
given by [cf.\ \eqref{e:post-mrc-snr}]
\begin{equation}
\SNR_\mathrm{MRC} = \sum_{m'=1}^{m} \SNR_{m',l}(\beta_m),
\label{e:tv-post-mrc-snr}
\end{equation}
with $\SNR_{m',l}(\beta_m)$ is as given in \eqref{e:tv-SNRml}.

\subsection*{Efficiency Analysis}

To show that the resulting scheme is asymptotically perfect, we first
note that when random dither encoding, MRC decoding, and
capacity-achieving base codes are used, the mutual information
$I'_{m,l}$ satisfies [cf.\ \eqref{e:eval}]
\begin{equation}
I'_{m,l} \geq \log\left(1+ \SNR_\mathrm{MRC}(\beta_m) \right) 
\label{e:tv-eval}
\end{equation}
with $\SNR_\mathrm{MRC}(\beta_m)$ as in \eqref{e:tv-post-mrc-snr}.

Again the efficiency of the scheme depends on the choice of power
allocation matrix \eqref{e:P}.  Recall that we may further bound
$I'_{m,l}$ for all $m$ by \eqref{e:sand}. Thus, if we choose the rate
$R''/L$ of the base code in each layer to be \eqref{e:conserve} then
\eqref{e:sand} ensures decodability after $m$ blocks are received when
the channel gain satisfies $|\beta_m| \geq \alpha_m$, as required.
Moreover, the efficiency $R''/R$ can be made as close as desired to
one by taking $L$ sufficiently large.

\section*{Acknowledgments}

We thank the reviewers and associate editor for their helpful
comments, which led to a number of improvements in the manuscript.  We
also acknowledge the efforts of Yuval Har-Zion, whose project at Tel
Aviv University provided the simulation results depicted in
Fig.~\ref{f:sim}.

\end{document}